\documentclass[prb,aps,showpacs,twocolumn,preprintnumbers,
amsmath,amssymb,superscriptaddress,longbibliography]{revtex4-2}
\usepackage[english]{babel}
\usepackage{amsmath,amssymb,amsfonts}
\usepackage{graphicx}
\usepackage[colorlinks=True,linkcolor=red,citecolor=blue,urlcolor=blue]{hyperref}

\usepackage{booktabs}
\usepackage[dvipsnames]{xcolor}
\usepackage{braket}
\usepackage{bm}
\usepackage{bbm}

\usepackage{enumitem}
\usepackage{verbatim}

\usepackage{braket}
\usepackage{float}
\usepackage{multirow}
\usepackage{longtable}
\usepackage[normalem]{ulem}
\usepackage{array}
\usepackage{makecell}
\usepackage{subfigure}
\usepackage{booktabs}
\usepackage{multirow}
\usepackage{graphicx}
\graphicspath{{NewFigures/}}
\usepackage{dcolumn}
\usepackage{bm}
\usepackage{longtable}
\usepackage{xcolor}
\newcolumntype{C}{>{$}c<{$}}

\usepackage{listings}
\usepackage{color}

\definecolor{dkgreen}{rgb}{0,0.6,0}
\definecolor{gray}{rgb}{0.5,0.5,0.5}
\definecolor{mauve}{rgb}{0.58,0,0.82}

\allowdisplaybreaks
\DeclareMathAlphabet{\zc}{OT1}{pzc}{m}{it}

\begin{document}  

\title{Exceptional activated mode theory for generalized real-complex transitions}

\author{Mengjie Yang}
\affiliation{Department of Physics, National University of Singapore, Singapore 117551, Singapore}

\author{Alexander N. Poddubny}
\email{poddubny@weizmann.ac.il}
\affiliation{Department of Physics of Complex Systems, Weizmann Institute of Science, Rehovot 7610001, Israel}

\author{Ching Hua Lee}
\email{phylch@nus.edu.sg}
\affiliation{Department of Physics, National University of Singapore, Singapore 117551, Singapore}

\date{\today}

\begin{abstract}
Real-to-complex spectral transitions mark the onset of amplification in non-Hermitian systems, but their thresholds are often treated as model-specific quantities. 
Here we develop a general, non-perturbative activated-mode principle that governs the real-to-complex threshold across broad classes of non-Hermitian systems.
A central insight is that only a small Hilbert subspace is ``activated" at the transition onset, which can be variationally determined through the competition between spectral detuning and mode-level projected non-Hermitian couplings. 
The result is a closed-form exceptional-activation condition for arbitrarily large ``disturbances", rather than a perturbative estimate. 
We apply our framework to three contrasting illustrative problems, establishing (i) a closed-form threshold for critical non-Hermitian skin amplification at \emph{all} system sizes; (ii) a new link between impurity tunneling threshold and exceptional point switching; and (iii) activation channel switching without underlying topological phase transition. Overall, our findings recast real-to-complex transitions as generic mode-selection problems independent of any specific symmetry.
\end{abstract}

\maketitle

Real-to-complex spectral transitions, where the eigenvalue spectrum $\{E\}$ of a non-Hermitian operator goes from purely real to partly complex, appear with striking universality across non-Hermitian platforms: $\mathcal{PT}$-symmetric photonics~\cite{Bender1998RealSpectra,Ruter2010ObservationPTOptics,Feng2017NonHermitianPhotonics,Hodaei2014MicroringLasers,Ozdemir2019PTPhotonics}, acoustic and mechanical metamaterials~\cite{Zhu2014PTAcoustics,Fleury2015InvisibleAcousticSensor,Brandenbourger2019NonReciprocalRobotic,Ghatak2020ActiveMechanical,Scheibner2020ActiveElastic}, topolectrical  circuits~\cite{Helbig2020Topolectrical,Hofmann2020ReciprocalSkin,sahin2025topolectrical}, dissipative cold-atom and trapped-ion setups~\cite{Li2019DissipativeFloquetAtoms,Lapp2019TunableLocalLoss,Ren2022ChiralControl,Liang2022DynamicSignatures}, and active, biological or disordered networks~\cite{Murugan2017BiologicalNetworks,Tang2021StochasticSystems,Yang2024Percolation}. Since $\mathrm{Im}\,E$ sets the exponential growth rate, the first such transition determines when a system begins to amplify, lase, or destabilize, making its threshold the sharpest dynamical fingerprint of the underlying non-Hermitian mechanism and the cleanest knob for its engineering~\cite{Ashida2020NonHermitianPhysics,Bergholtz2021ExceptionalTopology,ElGanainy2018NonHermitianPhysics,Heiss2012ExceptionalPoints,Miri2019ExceptionalPoints,meng2024exceptional}. 

Across these settings, a useful minimal description is often a real-spectrum reference problem assembled from coupled components: balanced gain-loss sublattices~\cite{Lieu2018NHSSH,Weimann2017PTCrystals}, non-Hermitian ladders that remain real in the decoupled limit~\cite{Li2020CriticalNHSE,Liu2020HelicalDamping,Yokomizo2022NonHermitianWaves,Qin2023UniversalCompetitive,qin2026anyon}, topological boundary modes coupled to bulk or defect channels~\cite{Malzard2015DefectStates,Leykam2017EdgeModes,Ni2018PTEdgeDomainWalls,yang2025beyond}, or structured reservoirs~\cite{Garmon2017CoalescingEigenvalues}. This common structure, however, has not yet led to a general threshold theory. Existing predictions remain model-by-model: perturbation theory~\cite{Brody2014BiorthogonalQM,Demange2012ThreeCoalescing,Mostafazadeh2010PseudoHermitian} fails beyond weak coupling; non-Bloch constructions~\cite{Yokomizo2019NonBloch,Yang2020GBZ,Yao2018EdgeStates,zhang2021tidal,zhang2022real,yang2022designing,yokomizo2023non,lei2024activating,meng2025generalized} assume translation-invariant thermodynamic limits; and transfer-matrix or Green-function analyses~\cite{Kunst2018BiorthogonalBBC,Okuma2020TopologicalOrigin,Borgnia2020BoundaryModes} are model-specific. A scaling formula for arbitrary component couplings is still missing.

\begin{figure*}[!htp]
    \centering
    \includegraphics[width=0.8\linewidth]{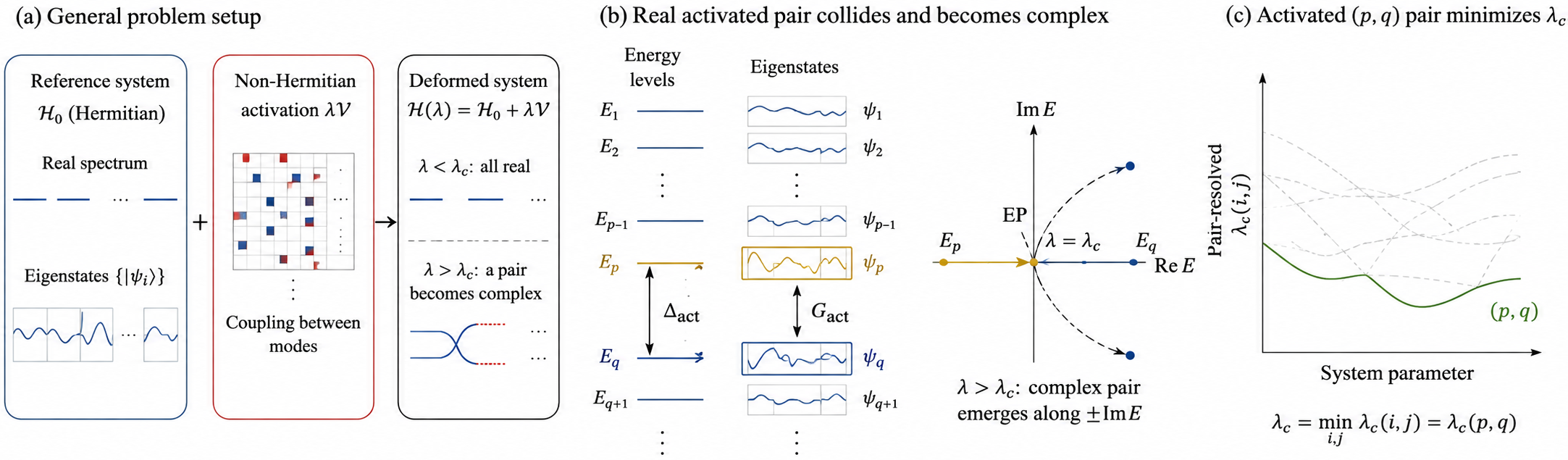}
    \caption{Activated-mode mechanism and real-to-complex threshold selection.
(a) Operator family setup $\mathcal{H}(\lambda)=\mathcal{H}_0+\lambda\mathcal{V}$: $\mathcal{H}_0$ has a real reference spectrum, while $\lambda\mathcal{V}$ activates complex eigenenergies. We seek the minimal coupling $\lambda=\lambda_c$ at which the first eigenenergies leave the real axis.
(b) Projecting $\mathcal V$ onto all candidate pairs identifies the activated pair $(p,q)$ with the lowest pair-resolved EP threshold $\lambda_c^{(i,j)}$. Its threshold $\lambda_c$ is set by the competition between the reference spectral detuning $\Delta_{\rm act}$ and the projected activation strength $G_{\rm act}$. With increasing $\lambda$, the selected levels approach along $\operatorname{Re}E$, coalesce at an EP, and subsequently escape into the complex plane.
(c) As any system parameter is varied, the pair-resolved thresholds $\lambda_c^{(i,j)}$ can cross, with activated mode pairs correspondingly switched. The physical threshold is their lower envelope,
$\lambda_c=\min_{i<j}\lambda_c^{(i,j)}=\lambda_c^{(p,q)}$
[Eq.~\eqref{eq:threshold_minimization}]. The EP condition is imposed on the active block itself, without a finite-order expansion in $\lambda$.
}
    \label{fig:fig1}
\end{figure*}

For this purpose, we develop a generic \emph{exceptional activated mode theory} framework. The central insight is that a generic
real-to-complex transition is governed by only small Hilbert-space sector: most commonly, a single ``activated" mode pair that collides at an exceptional point (EP). 
Projecting the control coupling onto all candidate pairs gives a pair-resolved transition threshold, and the physical transition is their lower envelope---turning the onset of amplification into a finite-size variational problem over the reference modes, characterized by 
the competition between the spectral detuning and projected non-Hermitian couplings. 
This yields a unified, non-perturbative expression for the transition scaling, valid across qualitatively different amplification mechanisms, as demonstrated through three contrasting examples.

\noindent \textit{Activated mode theory: exact transition thresholds beyond the perturbative regime.---} We consider a finite-dimensional one-parameter family of Hamiltonians 
\begin{equation} 
\mathcal{H}(\lambda) = \mathcal{H}_0+\lambda \mathcal{V}, \qquad \lambda\in\mathbb{R}. 
\label{eq:generic_problem} 
\end{equation} 
Here, $\mathcal H_0$ is diagonalizable with a real spectrum in the window of interest, while $\mathcal V$ is controlled by real $\lambda$, which may be large [Fig.~\ref{fig:fig1}(a)]. Neither $\mathcal{H}_0$ nor $\mathcal{V}$ is assumed to be Hermitian, local, translation-invariant, or protected by any specific symmetry. This work focuses on determining $\lambda_c$, the \emph{minimal} $\lambda$ for which the spectrum of $\mathcal{H}(\lambda)$ ceases to be real.

More concretely, as $\lambda$ is increased from zero, the eigenenergies of $\mathcal{H}(\lambda)$ will move around, initially along the real axis. Eventually, they will start colliding: if $\mathcal{H}(\lambda)$ is non-Hermitian, the first pair may collide at an EP and then be ``activated" to leave the real axis [Fig.~\ref{fig:fig1}(b)]. Our goal is to predict this threshold $\lambda=\lambda_c$ directly from the few 
relevant reference modes of $\mathcal{H}_0$, and the operator $\mathcal{V}$ projected onto their subspaces. 
Generically, this first collision EP is a two-branch coalescence; higher-order degeneracies correspond to remote-mode corrections as discussed in Supplemental Materials (SM)~S2~\cite{suppmat}.

We assume throughout that $\mathcal{H}_0$ is diagonalizable in the spectral window of interest. Let $\{|\psi_i^R\rangle,|\psi_i^L\rangle\}$ be its biorthogonal eigenmodes~\cite{Moiseyev2011NonHermitianQM,Brody2014BiorthogonalQM,Ashida2020NonHermitianPhysics}, with real energies $E_i$ and $\langle\psi_i^L|\psi_j^R\rangle=\delta_{ij}$. For a candidate pair $(i,j)$, its irreducible two-branch block takes the form 
\begin{equation}
    \mathcal{H}_{ij}^{\rm EP}(\lambda)
    =
    \begin{pmatrix}
    \tilde E_i(\lambda) & \lambda A_{ij}\\
    \lambda B_{ji} & \tilde E_j(\lambda)
    \end{pmatrix},
    \label{eq:active_block}
\end{equation} 
where $\tilde E_i=E_i+\lambda V_{ii}$, $A_{ij}=V_{ij}$ and $B_{ji}=V_{ji}$ depend on the projected matrix elements $V_{ij}=\langle\psi_i^L|\mathcal V|\psi_j^R\rangle$ of the activation operator.  
The EP activation condition is the vanishing of the discriminant of Eq.~\eqref{eq:active_block},
\begin{equation}
    \left(\Delta E_{ij}+\lambda\Delta D_{ij}\right)^2
    +
    4\lambda^2A_{ij}B_{ji}
    =0,
    \label{eq:discriminant_condition}
\end{equation}
where, in terms of the mode projector $P_i=|\psi_i^R\rangle\langle\psi_i^L|$,
\begin{equation}
\begin{aligned}
\Delta E_{ij}
&=E_i-E_j=\operatorname{Tr}\!\left[(P_i-P_j)\mathcal H_0\right],\\
\Delta D_{ij}
&=
\langle\psi_i^L|\mathcal V|\psi_i^R\rangle
-
\langle\psi_j^L|\mathcal V|\psi_j^R\rangle=\operatorname{Tr}\!\left[(P_i-P_j)\mathcal V\right],\\
\quad A_{ij}B_{ji}&=\operatorname{Tr}[P_i\mathcal VP_j\mathcal V].
\end{aligned}
\label{eq:defEAB}
\end{equation}
$A_{ij}B_{ji}=\operatorname{Tr}[P_i\mathcal VP_j\mathcal V]$ measures the biorthogonal weight of $\mathcal V$ on the pair; for local or sparse $\mathcal V$, it acts as a spatial filter.

For real projected matrix elements and $A_{ij}B_{ji}<0$, 
Eq.~\eqref{eq:discriminant_condition} gives the two candidate EP branches
\begin{equation}
    \lambda_{c,\sigma}^{(i,j)}
    =
    -\frac{\Delta E_{ij}}
    {\Delta D_{ij}-2\sigma\sqrt{-A_{ij}B_{ji}}},
    \qquad
    \sigma=\pm .
    \label{eq:two_branch_general}
\end{equation}
For a sweep toward positive $\lambda$, the threshold of channel $(i,j)$ is
the smaller positive branch,
$\lambda_c^{(i,j)}
=\min_{\sigma=\pm:\lambda_{c,\sigma}^{(i,j)}>0}
\lambda_{c,\sigma}^{(i,j)}$.

The activated channel across all $(i,j)$ that precipitates the physical real-to-complex transition 
is the minimizing candidate pair [Fig.~\ref{fig:fig1}(c)]. Denoting this pair by $(p,q)$, i.e.,
$
(p,q)\in\operatorname*{arg\,min}_{i<j}\lambda_c^{(i,j)},
$
the physical $\lambda$ threshold is
\begin{equation}
    \lambda_c
    =
    \min_{i<j}\lambda_c^{(i,j)}
    =
    \lambda_c^{(p,q)}.
\label{eq:threshold_minimization}
\end{equation}
For the activated channel $(p,q)$ and its physical branch $\sigma_{pq}$, the threshold
takes the transparent competition form
\begin{equation}
\lambda_c=\frac{\Delta_{\rm act}}{G_{\rm act}},
\quad
\begin{aligned}
\Delta_{\rm act}&=|\Delta E_{pq}|,\\
G_{\rm act}&=\big|\Delta D_{pq}-2\sigma_{pq}\sqrt{-A_{pq}B_{qp}}\big|,
\end{aligned}
\label{eq:act_ratio}
\end{equation}
i.e. the activated spectral detuning $\Delta_{\rm act}$ competes against the projected
activation strength $G_{\rm act}$ [Fig.~\ref{fig:fig1}(b)].
The pair-resolved thresholds are candidate EP conditions for different projected channels, and the physical transition is the lowest admissible threshold whose branches remain real over $0\leq\lambda<\lambda_c$; see End Matter Sec.~I.

Two model-independent rules follow. The sign condition $A_{ij}B_{ji}<0$ is necessary---a Hermitian off-diagonal coupling has $B_{ji}=A_{ij}^{*}$ and fails it---and, by Eq.~\eqref{eq:act_ratio}, the activated pair is selected not by the smallest reference detuning but by the smallest $\Delta_{\rm act}/G_{\rm act}$ ratio.

Eqs.~\eqref{eq:active_block}--\eqref{eq:two_branch_general} are not a finite-order expansion in $\lambda$, unlike ordinary perturbative treatments of non-Hermitian degeneracies~\cite{Kato1995,Demange2012ThreeCoalescing,Brody2014BiorthogonalQM}: once the active sector is identified, its EP condition is solved without expanding in $\lambda$, as verified against exact numerics up to $\lambda_c=\mathcal O(1)$ [see SM~\cite{suppmat}~S1] for our first illustrative example below.

\begin{figure}[t]
    \centering
    \includegraphics[width=0.98\linewidth]{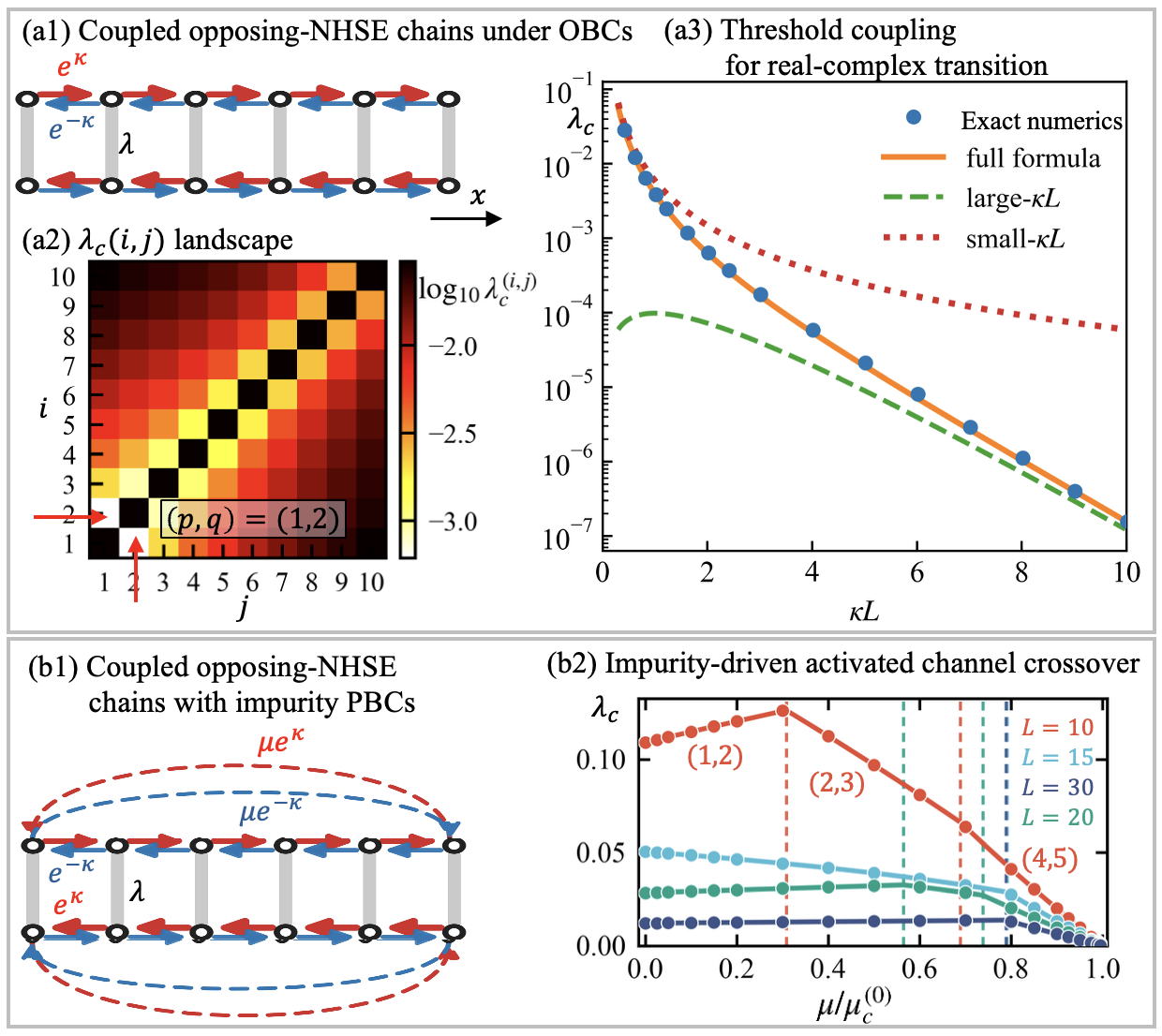}
    \caption{Boundary-controlled activation in antagonistic NHSE systems. Both (a,b) possess the same oppositely nonreciprocal HN bulk hoppings (inverse skin lengths $\pm\kappa$) and inter-chain couplings $\lambda$, but differ in boundary closure: row (a) keeps the chains open, while row (b) closes them by end-to-end impurity links $\mu e^{\pm\kappa}$. (a1) OBC realization [Eqs.~\eqref{eq:HN_model}--\eqref{eq:Vperp}]. (a2) Pair-resolved two-mode thresholds $\lambda_c^{(i,j)}$ [Eq.~\eqref{eq:HN_pair_threshold_compact}]; in the displayed $10\times10$ low-index window ($L=100$, $\kappa=0.02$) the lower envelope selects the band-edge channel $(p,q)=(1,2)$. (a3) $\lambda_c$ threshold, with exact numerics in excellent agreement with our new closed-form expression Eq.~\eqref{eq:HN_closed_threshold}, which connects the large-$\kappa L$ asymptote [Eq.~\eqref{eq:HN_large_kL}] to the small-$\kappa L$ one [Eq.~\eqref{eq:HN_small_kL}]---a crossover between the skin-overlap and algebraic band-edge regimes. (b1) Realization of the same construction, but with impurity-PBC closure of strength $\mu$ [Eqs.~\eqref{eq:SF_model} and \eqref{eq:Vperp}]. (b2) Boundary-driven $\lambda_c$ threshold as a function of $\mu/\mu_c^{(0)}$, where $\mu_c^{(0)}$ is the impurity threshold that drives the uncoupled ($\lambda=0$) reference chain spectrum complex. 
        Dots denote exact numerical data, which agree excellently with solid line segments representing theoretical predictions [Eq.~\eqref{eq:SF_pair_threshold}]. Vertical dashed lines mark the switching of the activated channels at various $L$. Beyond the rightmost switch, the $\lambda_c$ threshold decreases linearly toward zero as $\mu\rightarrow\mu_c^{(0)}$. 
        }
    \label{fig:cNHSE}
\end{figure}

\noindent\textit{Canonical cNHSE: Beyond the asymptotic exponential law.--}
This simplest warm-up example consists of two coupled antagonistic Hatano-Nelson (HN) chains~\cite{Hatano1996Localization,Hatano1997VortexDepinning} under open boundary conditions (OBCs) [Fig.~\ref{fig:cNHSE}(a)]. 
With oppositely-directed bond asymmetry and hence NHSE pumping, this coupled setup is the minimal setting for unconventional scaling due to NHSE feedback loops (i.e. critical NHSE or cNHSE), challenging 
conventional notions of non-Heritian bulk-boundary correspondences~\cite{Yao2018EdgeStates,Kunst2018BiorthogonalBBC,Yokomizo2019NonBloch,Okuma2020TopologicalOrigin,Zhang2020Winding,Yang2020GBZ}. We write the model as
\begin{equation}
    \mathcal{H}_{\text{coupled-HN}}(\lambda)
    =
    \mathcal{H}_{0}
    +
    \lambda\mathcal{V}_{\perp},
    \label{eq:HN_model}
\end{equation}
where the uncoupled chains form the reference operator
\begin{equation}
    \mathcal{H}_{0}
    =
    \sum_{x=1}^{L-1}
    [
    e^{\kappa}a_{x+1}^{\dagger}a_x^{\vphantom{\dagger}}
    +
    e^{-\kappa}a_x^{\dagger}a_{x+1}^{\vphantom{\dagger}}
    +
    e^{-\kappa}b_{x+1}^{\dagger}b_x^{\vphantom{\dagger}}
    +
    e^{\kappa}b_x^{\dagger}b_{x+1}^{\vphantom{\dagger}}
    ],
    \label{eq:HN_K0}
\end{equation}
as shown in Fig.~\ref{fig:cNHSE}(a1).
Its spectrum stays real because the gauge transformation $a_x\to e^{-\kappa x}a_x$, $b_x\to e^{\kappa x}b_x$ removes the asymmetric hopping from both chains, leaving an equivalent Hermitian nearest-neighbor chain. The inter-chain coupling is $\lambda \mathcal{V}_{\perp}$, where
\begin{equation}
    \mathcal{V}_{\perp}
    =
    \sum_{x=1}^{L}
    \left(
    a_x^{\dagger}b_x^{\vphantom{\dagger}}
    +
    b_x^{\dagger}a_x^{\vphantom{\dagger}}
    \right).
    \label{eq:Vperp}
\end{equation}
In the strong NHSE asymptotic limit of $\kappa L \gg 1$~\cite{Li2020CriticalNHSE,Qin2023UniversalCompetitive}, the conventional cNHSE argument gives a exponentially weak $\lambda_c\sim e^{-\kappa L}$ threshold, 
as elaborated in End Matter Sec.~II. 
Here below, however, we shall derive the full scaling behavior of $\lambda_c$ and identify the activated pair controlling it, up to the strong $\lambda_c$ (i.e. $\kappa L\ll 1$) regime where the NHSE clearly does not dominate.

Projecting the inter-chain coupling onto the reference modes gives the pair-resolved thresholds $\lambda_c^{(i,j)}=(L+1)e^{-\kappa(L+1)}\delta_{ij}/W_{ij}$, with $\delta_{ij}=\cos\frac{i\pi}{L+1}-\cos\frac{j\pi}{L+1}$ and the biorthogonal overlap weight $W_{ij}$ given in End Matter Sec.~III and SM~\cite{suppmat}~S3. Minimizing over all pairs selects the band-edge channel $(p,q)=(1,2)$ as the activated one [Fig.~\ref{fig:cNHSE}(a2)], from which the activated-mode threshold takes the closed-form expression 
\begin{equation}
\begin{aligned}
\lambda_c
=\frac{4(L+1)e^{-\kappa(L+1)}}{Q}
\sin\frac{3\pi}{2(L+1)}
\sin\frac{\pi}{2(L+1)},
\end{aligned}
\label{eq:HN_closed_threshold}
\end{equation}
where $Q=2C_0+2C_1-C_2-2C_3-C_4$ and $C_m=\sum_{x=1}^{L}e^{-2\kappa x}\cos\frac{m\pi x}{L+1}.$
Eq.~\eqref{eq:HN_closed_threshold} determines the entire crossover, in perfect agreement with numerics [Fig.~\ref{fig:cNHSE}(a3)], without any $\lambda$ expansion or thermodynamic-limit approximation.
In the strong and weak NHSE limits, Eq.~\eqref{eq:HN_closed_threshold} reduces to
\begin{align}
\lambda_c&\simeq \tfrac23(L+1)\sinh^3\kappa\,
\operatorname{sech}\kappa\,e^{-\kappa(L+1)}
&&\kappa L\gg1, \label{eq:HN_large_kL}\\ \lambda_c &\simeq 3\pi^2\Big/[2(L+1)^2], &&\kappa L\ll1.
\label{eq:HN_small_kL}
\end{align}
The strong-NHSE limit approximately reduces to a logarithmic $L$-scaling law for fixed inter-chain coupling, whereas the small-$\kappa L$ limit applies when the skin depth exceeds the system size.
The agreement between the exact threshold (blue dots) and Eq.~\eqref{eq:HN_closed_threshold} is excellent across the full crossover, confirming that the real-to-complex onset is indeed controlled by the selected $(1,2)$ active channel. The large-$\kappa L$ and small-$\kappa L$ asymptotes reproduce the corresponding ends of the finite-size curve, showing that the same expression connects the conventional exponentially small skin-overlap threshold to the algebraic band-edge threshold. 

This is the part missed by the usual asymptotic cNHSE estimate~\cite{Li2020CriticalNHSE,Qin2023UniversalCompetitive,Yokomizo2022NonHermitianWaves,Zhang2022Universal,Wang2023LongRange}. The conventional argument correctly identifies the exponentially small scale generated by opposite skin accumulation, but it does not determine the finite-size crossover or the active pair responsible for the first instability. Equation~\eqref{eq:HN_closed_threshold} supplies both pieces of information. Surprisingly, the exponential skin-overlap and algebraic band-edge regimes ($\kappa L \gg 1$ and $\kappa L \ll 1$ respectively) are not controlled by different mechanisms or different mode pairs: both originate from the same exceptional activated channel $(1,2)$, and are continuously connected by the same hitherto un-appreciated unifying mechanism despite their qualitatively different scaling behavior.

\noindent\textit{Impurity cNHSE: boundary closure controls the activation and scaling.--}
We next close the open ends of the coupled HN ladder in Eq.~\eqref{eq:HN_model} by end-to-end impurity hoppings [dashed curves in Fig.~\ref{fig:cNHSE}(b1)]. Although confined to the boundary, these bonds alter both the critical scaling of the full ladder and the identity of the activated channel, providing a boundary-controlled interplay between impurity feedback and NHSE competition. Keeping all bulk couplings the same as before, the Hamiltonian is
\begin{equation}
\mathcal H_{\rm imp}(\lambda,\mu)
=
\mathcal H_0+\mathcal B_\mu+\lambda\mathcal V_{\perp},
\label{eq:SF_model}
\end{equation}
where the extra end-to-end impurity is described by
\begin{equation}
\begin{aligned}
\mathcal B_\mu
=\mu(
&e^{\vphantom{-}\kappa}a_{1}^{\dagger}a_L
+e^{-\kappa}a_L^{\dagger}a_1
+e^{-\kappa}b_{1}^{\dagger}b_L
+e^{\vphantom{-}\kappa}b_L^{\dagger}b_1
).
\end{aligned}
\label{eq:SF_boundary}
\end{equation}
The open cNHSE ladder is recovered at $\mu=0$, while finite $\mu$ enables additional PBC-like tunneling that complements the cNHSE amplification, potentially hastening the real-to-complex transition. This physics is distinct from the impurity-induced scale-free behavior in each uncoupled chain~\cite{Li2021SF,Guo2021GBC}, which occurs only well \emph{after} the real-to-complex transition.

For each reference single chain, the boundary constraints [see SM~\cite{suppmat}~Sec.~S4B] shift the ``momenta" $k_n=k_n(\mu)$ from open waveguide ($\mu=0$) quantized values of $k_n(0)=n\pi/(L+1)$ to the solutions of
\begin{equation}
\begin{aligned}
&\sin[(L+1)k_n]-\mu^{2}\sin[(L-1)k_n]-2\mu\cosh(\kappa L)\sin k_n=0,
\end{aligned}
\label{eq:EM_SF_secular}
\end{equation}
with reference energies $E_n(\mu)=2\cos k_n(\mu)$. Besides shifting $k_n$, the boundary coupling $\mu$ also introduce a boundary-reflected wave of relative weight $\mu e^{-\eta_\ell\kappa L}$, giving rise to right eigenmode profiles  $\langle x|\psi^{R}_{n,\ell}(\mu)\rangle\propto $
\begin{equation}
\begin{aligned}
e^{\eta_\ell\kappa(x-1)}
\Big[\sin\big((L{+}1{-}x)k_n(\mu)\big)
+\mu e^{-\eta_\ell\kappa L}
\sin\big((x{-}1)k_n(\mu)\big)\Big],
\end{aligned}
\label{eq:EM_SF_reference_modes}
\end{equation}
where $\eta_{a,b}=\pm1.$ Through these spatial profiles, $\mu$ controls the $\lambda_c$ threshold by modifying the reference detuning $E_i(\mu)-E_j(\mu)$ and the projected rung elements $v_{mn}(\mu)$ [see SM~\cite{suppmat}, Secs.~S4B and S4D], based on the \emph{same} EP condition used at $\mu=0$.

Although the boundary impurity links $\mu e^{\pm\kappa}$ invalidate the non-unitary gauge transformation that guarantees real spectrum in each reference chain, the reference operator $\mathcal H_0+\mathcal B_\mu$ still retains a real spectrum for sufficiently small $\mu$. Increasing $\mu$ eventually produces an isolated-chain exceptional point at $\mu=\mu_c^{(0)}$. For $L=4\ell$ with $\ell\in\mathbb N$, the first coalescence can be shown to occur at $k_c=\pi/2$, yielding, at $\lambda=0$, the nice \emph{impurity} threshold
\begin{equation}
\mu_c^{(0)}|_{L=4l}=e^{-\kappa L}.
\label{eq:SF_reference_EP}
\end{equation}
For generic $L$, $\mu_c^{(0)}$ is determined by the coalescence of two real roots of the exact boundary equation (End Matter Sec.~IV). The previously discussed positive $\lambda$-driven threshold therefore only exists for $0\leq\mu<\mu_c^{(0)}$, with $\lambda_c(\mu)=0$ for $\mu\geq\mu_c^{(0)}$.

Ordering the reference energies as $E_1(\mu)>\cdots>E_L(\mu)$ and using paired biorthogonal normalization, define
$
v_{mn}(\mu)
=
\langle\psi_{m,b}^{L}(\mu)|
\mathcal V_{\perp}
|\psi_{n,a}^{R}(\mu)\rangle
=
\langle\psi_{m,a}^{L}(\mu)|
\mathcal V_{\perp}
|\psi_{n,b}^{R}(\mu)\rangle
$
for $m,n\in\{i,j\}$.
As in the OBC ladder, equal and opposite chain combinations reduce the two-level projection to the same $2\times2$ active blocks [see Eq.~\eqref{eq:HN_chain_parity_blocks}]. For $i<j$ and $v_{ij}(\mu)v_{ji}(\mu)<0$ the smallest positive threshold is 
\begin{equation}
    \lambda_c^{(i,j)}(\mu)=[E_i(\mu)-E_j(\mu)]/\Gamma_{ij}(\mu),\label{eq:SF_pair_threshold}
\end{equation} with $\Gamma_{ij}(\mu)=|v_{ii}(\mu)-v_{jj}(\mu)|+2\sqrt{-v_{ij}(\mu)v_{ji}(\mu)}$.
Thus, impurity boundary closure modifies both the spectral detuning and the projected activation strength within the same EP condition used under OBCs.

Applying the general minimization rule from Eq.~\eqref{eq:threshold_minimization}, the activated eigenenergy pair now changes as $\mu$ varies. Branch-resolved spectra [SM~\cite{suppmat}~S4, Figs.~S3--S4] confirm that this is a genuine switching of first-coalescence channel, not a relabeling of continuously evolving levels. Depending on $L$ and $\kappa$, the threshold may exhibit no switch, a direct switch, or several intermediate activated channels. The agreement between the solid curves and exact numerics in Fig.~\ref{fig:cNHSE}(b2) confirms this channel-selection mechanism.

Notably, the $L=10$ threshold (red) in Fig.~\ref{fig:cNHSE}(b2) is nonmonotonic in $\mu$, suggestive of distinct competing physical mechanisms behind the EP activation. At small $\mu$, the activated channel remains $(1,2)$, and the boundary closure pushes $E_1$ up and $E_2$ down, widening the detuning and raising $\lambda_c$ linearly in $\mu$ [SM~\cite{suppmat}~S4]. As $\mu$ grows the minimizing channel switches, $(1,2)\to(2,3)\to(4,5)$, the terminal pair being the one that has to coalesce at the isolated-chain EP: there $\Delta E_{45}\sim(\mu_c^{(0)}-\mu)^{1/2}$ while $\Gamma_{45}\sim(\mu_c^{(0)}-\mu)^{-1/2}$, so $\lambda_c$ collapses linearly to zero. Different activated pairs therefore control different $\mu$ windows: at fixed $\lambda=0.115$ for instance, increasing $\mu$ curiously drives the spectrum first from complex to real, and then back to complex subsequently [SM~\cite{suppmat}~S4].

\begin{figure}[t]
    \centering
    \includegraphics[width=0.9\linewidth]{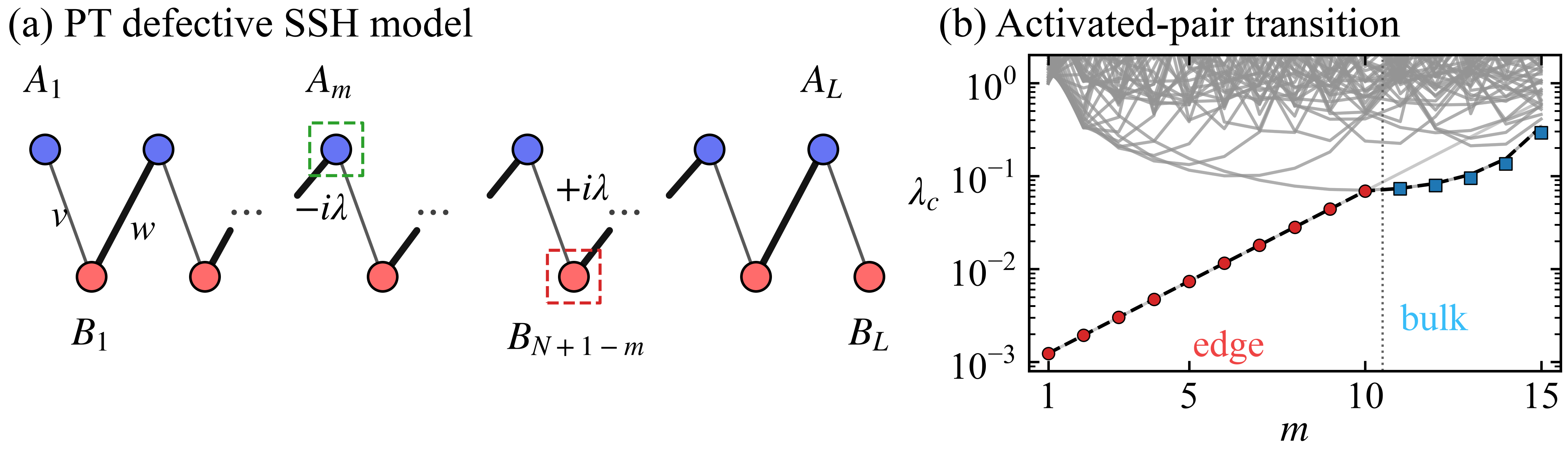}
   \caption{Seemingly non-local active-channel switching in a $\mathcal{PT}$-defective SSH chain. (a) OBC SSH chain with balanced gain and loss at $A_m$ and $B_{L+1-m}$ [Eqs.~\eqref{eq:SSH_model} and \eqref{eq:SSH_defect}]. (b) Real-to-complex transition threshold $\lambda_c$ versus $m$. Translucent gray curves show all pair-resolved candidates $\lambda_c^{(i,j)}(m)=|E_i-E_j|/|2A_{ij}|$; their lower envelope gives the active-space prediction (black dashed curve), agreeing perfectly with exact numerics (Red and Blue for edge- and bulk-activated channels, respectively). Their switching (vertical dotted line) occurs not only far from the topological edge $(m=1)$, but also unaccompanied by any bulk topological transition; in the edge regime, $\lambda_c$ follows Eq.~\eqref{eq:SSH_top_closed}. 
   }
    \label{fig:ssh}
\end{figure}

\noindent\textit{Ostensibly non-local topological-to-bulk mode activation switching.--}
We next show that the activated exceptional mechanism extends beyond coupled non-Hermitian systems, with novel physical implications, and that $\mathcal{V}_m$ does not have to represent physical couplings. Here, our reference operator $\mathcal{H}_0$ is a Hermitian topological SSH chain~\cite{Su1979Solitons,Heeger1988Solitons,Asboth2016ShortCourse}, while non-Hermiticity is introduced solely through the tunable activation operator $\lambda\mathcal{V}_m$, which assigns localized gain and loss to sites $A_m$ and $B_{L+1-m}$ related by reflection [Fig.~\ref{fig:ssh}(a)]:
\begin{equation}
    \mathcal{H}_{\rm SSH}(\lambda,m)=\mathcal{H}_0+\lambda\mathcal{V}_m,
    \label{eq:SSH_model}
\end{equation}
where $\mathcal H_0=\sum_x v(a_x^\dagger b_x+\mathrm{h.c.})+\sum_x w(a_{x+1}^\dagger b_x+\mathrm{h.c.})$ is the usual Hermitian SSH model with $|v|<|w|$, and the onsite gain and loss, with positions controlled by $m$, are implemented by
\begin{equation}
    \mathcal{V}_m=-i|A_m\rangle\langle A_m|+i|B_{L+1-m}\rangle\langle B_{L+1-m}|
    \label{eq:SSH_defect}
\end{equation}
Our model differs from previous non-Hermitian SSH constructions that use sublattice-wide gain and loss, non-Hermitian hoppings, or an engineered dimerization interface~\cite{Schomerus2013TopologicallyProtected,Lieu2018NHSSH,Weimann2017PTCrystals}; the closest precedent is a reflection-symmetric pair of gain--loss defects~\cite{Jin2017PTDefects}. We use that geometry for a different purpose: with $\mathcal H_0$ and its bulk topology fixed, we resolve the real-to-complex transition into competing pair-resolved EP channels, and ask which one the position control knob $m$ selects. 
The bulk invariant guarantees topological edge modes but does not decide whether the edge pair or a bulk pair gives the lowest threshold; that is settled by Eq.~\eqref{eq:threshold_minimization} alone.

Since $\mathcal{H}_0$ is Hermitian, $|\psi_i^R\rangle=|\psi_i^L\rangle\equiv|\psi_i\rangle$. For a candidate pair $(i,j)$, the projected non-Hermitian coupling onto it is characterized by the off-diagonal matrix elements
$
A_{ij}=\langle\psi_i|\mathcal V_m|\psi_j\rangle
$
and
$
B_{ji}=\langle\psi_j|\mathcal V_m|\psi_i\rangle
$,
as defined generally in Eq.~\eqref{eq:defEAB}. Here $A_{ij}$ and $B_{ji}$ denote projected matrix elements,   not to be confused with the sublattice-site labels $A_m$ and $B_{L+1-m}$. For the balanced gain-loss profile in Eq.~\eqref{eq:SSH_defect}, they satisfy $A_{ij}=B_{ji}=
i\left[
\psi_i(A_m)\psi_j(A_m)
-
\psi_i(B_{L+1-m})\psi_j(B_{L+1-m})
\right].$
For opposite-parity SSH eigenstates, the diagonal difference vanishes, $\Delta D_{ij}=0$, and $A_{ij}B_{ji}<0$ whenever the projected defect matrix element is nonzero. Hence Eq.~\eqref{eq:two_branch_general} gives $\lambda_c^{(i,j)}(m)=|E_i-E_j|/|2A_{ij}|$. Figure~\ref{fig:ssh}(b) shows these pair-resolved thresholds as thick translucent gray curves. Their lower envelope, Eq.~\eqref{eq:threshold_minimization}, gives the physical threshold and conspicuously switches from the topological edge pair to a bulk pair as $m$ varies.
This showcases an edge-to-bulk activation transition that occurs without any change in topological invariant, correctly predicted only by comparing projected EP thresholds across all candidate pairs.

For the topological channel, characterized by $r=|v/w|<1$ with a well-localized, opposite-parity in-gap edge pair that has nonzero defect overlap and gives the global minimum in Eq.~\eqref{eq:threshold_minimization}, this threshold further simplifies to 
\begin{equation}
    \lambda_c^{\rm SSH}(m)\simeq |w|r^{L-2m+2}.
    \label{eq:SSH_top_closed}
\end{equation}
Saliently, as shown by the dashed black curve in Figure~\ref{fig:ssh}(b), the edge pair remains as the activated channel well beyond its localization core, till $m>10$ unit cells away from the boundary. 
This enigmatic observation is not due to non-local couplings---both $\mathcal H_0$ and $\mathcal V_m$ are strictly local. Instead, as elaborated in SM~\cite{suppmat}~Sec.~S5, this seemingly non-local influence of the topological edge state occurs because our EP activation mechanism involves the ratio of \emph{two} exponentially small quantities: the finite-size edge splitting $\propto r^L$, and the projected defect coupling $\propto r^{2m-2}$. 

\noindent\textit{Discussion.---}
We have developed a general EP-activation framework for predicting the precise mechanism behind real-to-complex transitions, with near-perfect agreement with exact numerics in our demonstrated examples. As a generic approach, it allows the prediction of new critical scaling laws with regards to different system sizes~\cite{Hatano1996Localization,Hatano1997VortexDepinning,Li2020CriticalNHSE,Zhang2022Universal,Wang2023LongRange,Yang2024Percolation,Yang2026EBB,yang2026reversing}, boundary conditions~\cite{Kunst2018BiorthogonalBBC,Yao2018EdgeStates,Yokomizo2019NonBloch,Okuma2020TopologicalOrigin,Zhang2020Winding,Yang2020GBZ,Guo2021GBC,yang2026reversing}, defect or impurity positions~\cite{Li2021SF,Bhargava2021Dislocation,Schindler2021Dislocation}, or the spatial profile and range of the activating channels~\cite{Yi2020Onsite,Li2020CriticalNHSE,Wang2023LongRange,Zhang2022NHSEReview,Lin2023TopologicalNHSE,Yang2026EBB,li2025phase}.
Being platform-independent, our results can be tested in photonic lattices, microring arrays, and quantum walks~\cite{Ruter2010ObservationPTOptics,Hodaei2014MicroringLasers,Weimann2017PTCrystals,Weidemann2020,Xiao2020,Xiao2021NonBlochPT}, acoustic and mechanical metamaterials~\cite{Zhu2014PTAcoustics,Fleury2015InvisibleAcousticSensor,Brandenbourger2019NonReciprocalRobotic,Ghatak2020ActiveMechanical,Scheibner2020ActiveElastic,Zhang2021Acoustic,Zhang2021HigherOrder,Wang2022MechanicalMorphing,Li2024DynamicMechanical,wang2025observation}, topolectrical circuits~\cite{Helbig2020Topolectrical,Hofmann2020ReciprocalSkin,Zou2021Circuit,Liu2021Circuit,rafi2022unconventional,Zhang2023CircuitSwitch,zou2024experimental}, and ultracold-atom platforms with controllable dissipation and synthetic dimensions~\cite{Li2019DissipativeFloquetAtoms,Lapp2019TunableLocalLoss,Ren2022ChiralControl,Liang2022DynamicSignatures,Zhou2022Ultracold,qin2024kinked,Zhao2025Ultracold}. 
Because the threshold is a \emph{ratio} [Eq.~\eqref{eq:act_ratio}], distinct microscopic
scalings of $\Delta_{\rm act}$ and $G_{\rm act}$ yield the same threshold law whenever
their exponent differences coincide (End Matter Sec.~IV).

\clearpage
\onecolumngrid
\section*{End Matter}

\twocolumngrid

\subsection*{Section I. Admissibility conditions and validity of the two-mode reduction}

For $\mathcal H(\lambda)=\mathcal H_0+\lambda\mathcal V$ [Eq.~\eqref{eq:generic_problem}],
a candidate pair $(i,j)$ of biorthogonal reference modes of $\mathcal H_0$ defines the
active block and its EP branches [Eqs.~\eqref{eq:active_block}--\eqref{eq:two_branch_general}]
\begin{equation}
\mathcal H_{ij}^{\rm EP}
=\!
\begin{pmatrix}
E_i+\lambda V_{ii} & \lambda A_{ij}\\
\lambda B_{ji} & E_j+\lambda V_{jj}
\end{pmatrix}\!,
\
\lambda_{c,\sigma}^{(i,j)}
=\frac{-\Delta E_{ij}}{\Delta D_{ij}-2\sigma\sqrt{-A_{ij}B_{ji}}},
\label{eq:EM1_block}
\end{equation}
with $V_{mn}=\langle\psi_m^L|\mathcal V|\psi_n^R\rangle$, $A_{ij}=V_{ij}$, $B_{ji}=V_{ji}$,
and, with $P_i=|\psi_i^R\rangle\langle\psi_i^L|$,
$\Delta E_{ij}=\operatorname{Tr}[(P_i-P_j)\mathcal H_0]$,
$\Delta D_{ij}=\operatorname{Tr}[(P_i-P_j)\mathcal V]$,
$A_{ij}B_{ji}=\operatorname{Tr}(P_i\mathcal VP_j\mathcal V)$ [Eq.~\eqref{eq:defEAB}].
Three conditions make Eq.~\eqref{eq:EM1_block} a physical transition.
\emph{(i) Sign}: $A_{ij}B_{ji}<0$ and real; this is impossible in Hermitian settings,
where $B_{ji}=A_{ij}^{\ast}$ and $A_{ij}B_{ji}=|A_{ij}|^2\ge0$, while genuinely complex
$A_{ij}B_{ji}$ admits a formal solution but a real positive threshold only under extra
tuning. \emph{(ii) Sweep direction}: only positive roots are retained, the negative ones
describing the opposite sweep, or equivalently $\mathcal V\to-\mathcal V$.
\emph{(iii) Spectral adjacency}: below an isolated second-order EP the two coalescing
branches must be adjacent in the ordered reference spectrum, since an intervening real
branch would be forced toward the same limiting eigenenergy, signaling a
higher-dimensional degeneracy; adjacency in turn imposes the parity restriction $i+j$
odd used in SM~\cite{suppmat}~S1.

Equation~\eqref{eq:EM1_block} is the \emph{direct} projection $P\mathcal H(\lambda)P$
with $P=P_p+P_q$, and thus discards the \emph{remote modes}, i.e. the reference modes
$r\notin\{p,q\}$ spectrally well separated from the activation energy $E_\ast$. Their
leading effect is the $O(\lambda^{2})$ Feshbach (L\"owdin) downfolding
$\lambda^2P\mathcal VQ(E-Q\mathcal H_0Q)^{-1}Q\mathcal VP$ with $Q=\mathbf 1-P$, which
shifts the entries of Eq.~\eqref{eq:EM1_block} by
$\delta\mathcal H_{ij}=\lambda^{2}\sum_{r\notin\{p,q\}}A_{ir}B_{rj}/(E_\ast-E_r)$ and
renormalizes both the detuning and $A_{pq},B_{qp}$ themselves. Writing
$C_{pq}=\max_{i,j}|\sum_{r}A_{ir}B_{rj}/(E_\ast-E_r)|$ and letting $s_{pq}$ be the
common scale of the retained entries $|\Delta E_{pq}|$, $|\lambda_c\Delta D_{pq}|$ and
$\lambda_c\sqrt{-A_{pq}B_{qp}}$ at the EP, the two-mode reduction is controlled when
\begin{equation}
\frac{\|\delta\mathcal H(E_\ast,\lambda_c)\|}{s_{pq}}
\sim
\lambda_c\,\frac{C_{pq}}{\sqrt{-A_{pq}B_{qp}}}\ll1 .
\label{eq:EM1_fractional}
\end{equation}
The admixture is thus parametrically subleading whenever $\lambda_c$ is itself small,
as it automatically is in both regimes below, where $\lambda_c$ is exponentially or
algebraically small in $L$. The admissibility conditions (i)--(iii), together with the
expected size of the remote-mode correction in Eq.~\eqref{eq:EM1_fractional}, can be
assessed directly from the reference data $\{E_i,P_i,\mathcal V\}$; when this correction
is not parametrically small, the projected prediction can instead be validated against
the full spectrum [SM~\cite{suppmat}~S1--S2]. Their violation is not a failure of the
framework, but a diagnostic that the activated sector exceeds two dimensions.

\subsection*{Section II. Origin and limitations of the conventional cNHSE scaling law}

The antagonistically coupled Hatano--Nelson (HN) ladder of
Eqs.~\eqref{eq:HN_model}--\eqref{eq:Vperp} scales unconventionally because the algebraic
structure of its characteristic equation changes singularly with the inter-chain
coupling. In two-chain block form,
$\mathcal H_{\text{coupled-HN}}=
\left(\begin{smallmatrix}H_b&\lambda\mathbbm 1_L\\ \lambda\mathbbm 1_L&H_a\end{smallmatrix}\right)$
with $H_{a,b}$ the open chains of Eq.~\eqref{eq:HN_K0} carrying inverse skin lengths
$\pm\kappa$, the off-diagonal blocks commute with everything, so that \emph{exactly}
\begin{equation}
\det\!\left[\mathcal H_{\text{coupled-HN}}(\lambda)-E\right]
=
\det\!\left[(H_b-E)(H_a-E)-\lambda^2\mathbbm 1_L\right].
\label{eq:EM2_det}
\end{equation}
At $\lambda=0$ this factorizes into $\det(H_b-E)\det(H_a-E)$, i.e. two decoupled chains,
whereas any $\lambda\neq0$ renders it generically irreducible: the zero-coupling and
thermodynamic limits do not commute, and a finite OBC system crosses over between the
decoupled and coupled thermodynamic spectra~\cite{Li2020CriticalNHSE}. The conventional
estimate then balances a feedback loop of two inter-chain tunnelings ($\lambda^2$)
against two NHSE amplification runs along the oppositely oriented chains
($e^{2\kappa L}$), i.e. $\lambda_c^2e^{2\kappa L}\sim O(1)$ and $\lambda_c\sim e^{-\kappa L}$.
This heuristic presupposes well-defined exponential skin amplification and correctly
gives the inverse-exponential law for $\kappa L\gg1$, but determines neither the
prefactor, nor the crossover once the skin depth $\kappa^{-1}$ exceeds $L$ (related
mesoscopic regimes arise in photonic settings~\cite{Sasha2024Mesoscopic}), nor which
mode pair drives the first instability. All three are supplied by
Eq.~\eqref{eq:HN_closed_threshold}, whose limits
Eqs.~\eqref{eq:HN_large_kL}--\eqref{eq:HN_small_kL} are generated by the \emph{same}
selected pair $(1,2)$.

\subsection*{Section III. Pair-resolved thresholds for the cNHSE ladder under OBCs}

Under OBCs, the gauge transformation $a_x\to e^{-\kappa x}a_x$, $b_x\to e^{\kappa x}b_x$
maps each chain of Eq.~\eqref{eq:HN_K0} onto the same reciprocal open chain, giving the
reference data $E_n=2\cos\frac{n\pi}{L+1}$ and
$\langle x|\psi_{n,\ell}^{R}\rangle\propto e^{\eta_\ell\kappa(x-1)}\sin\frac{n\pi x}{L+1}$
with $\eta_{a,b}=\pm1$, the left modes being their biorthogonal duals. Since each
spatial level appears once per chain, keeping the \emph{same} pair $(i,j)$ on
\emph{both} chains gives a four- rather than two-dimensional active space,
$\left(\begin{smallmatrix}\mathsf E_{ij}&\lambda\mathsf V_{ij}\\
\lambda\mathsf V_{ij}&\mathsf E_{ij}\end{smallmatrix}\right)$ with
$\mathsf E_{ij}={\rm diag}(E_i,E_j)$ and $(\mathsf V_{ij})_{mn}=v_{mn}$, which the
symmetric and antisymmetric chain combinations block-diagonalize exactly into
\begin{equation}
\mathcal M_{ij}^{(s)}(\lambda)
=
\begin{pmatrix}
E_i+s\lambda v_{ii} & s\lambda v_{ij}\\
s\lambda v_{ji} & E_j+s\lambda v_{jj}
\end{pmatrix},
\quad
s=\pm1,
\label{eq:HN_chain_parity_blocks}
\end{equation}
where $v_{mn}=\langle\psi_{m,b}^{L}|\mathcal V_{\perp}|\psi_{n,a}^{R}\rangle
=\langle\psi_{m,a}^{L}|\mathcal V_{\perp}|\psi_{n,b}^{R}\rangle$ under the paired
biorthogonal normalization. This four-to-two reduction uses only matrix elements of the
physical inter-chain coupling and therefore holds for both closures of the main text,
Eqs.~\eqref{eq:HN_model} and~\eqref{eq:SF_model}, which change only $E_n$ and $v_{mn}$.
Projecting $\mathcal V_\perp$ with the exact OBC skin modes above gives~\cite{suppmat}
$\Delta D_{ij}=-\frac{2e^{\kappa(L+1)}}{L+1}\sum_{x}e^{-2\kappa x}
[\sin^2\frac{i\pi x}{L+1}+\sin^2\frac{j\pi x}{L+1}]$ and
$A_{ij}=\frac{2e^{\kappa(L+1)}}{L+1}\sum_{x}e^{-2\kappa x}\sin\frac{i\pi x}{L+1}
\sin\frac{j\pi x}{L+1}=-B_{ji}$, so that $A_{ij}B_{ji}=-A_{ij}^2<0$ and condition (i)
of Sec.~I holds automatically for every pair with nonvanishing projected coupling.
Inserting these into Eq.~\eqref{eq:EM1_block} yields
\begin{equation}
\begin{aligned}
\lambda_c^{(i,j)}
&=(L+1)e^{-\kappa(L+1)}
\frac{\displaystyle
\cos\frac{i\pi}{L+1}-\cos\frac{j\pi}{L+1}}
{W_{ij}},\\
W_{ij}
&=\sum_{x=1}^{L}e^{-2\kappa x}
\left[
\sin^2\frac{i\pi x}{L+1}
+\sin^2\frac{j\pi x}{L+1}
\right]\\
&\quad+2\left|
\sum_{x=1}^{L}e^{-2\kappa x}
\sin\frac{i\pi x}{L+1}
\sin\frac{j\pi x}{L+1}
\right| ,
\end{aligned}
\label{eq:HN_pair_threshold_compact}
\end{equation}
whose minimization [Eq.~\eqref{eq:threshold_minimization}] selects the band-edge channel
$(p,q)=(1,2)$, degenerate with its spectral-reflection partner $(L-1,L)$, and yields
Eq.~\eqref{eq:HN_closed_threshold}; see SM~\cite{suppmat}~S3 for the selection proof.

\subsection*{Section IV. The impurity-closed reference operator and threshold collapse near a reference EP}

Closing the ladder with the impurity bonds
$\mathcal B_\mu=\mu(e^{\kappa}a_1^\dagger a_L+e^{-\kappa}a_L^\dagger a_1
+e^{-\kappa}b_1^\dagger b_L+e^{\kappa}b_L^\dagger b_1)$ [Eq.~\eqref{eq:SF_boundary}]
gives the reference operator $\mathcal H_0+\mathcal B_\mu$ of
Eq.~\eqref{eq:SF_model}. Each chain retains a real spectrum
$E_n(\mu)=2\cos k_n(\mu)$ as long as all physical roots of
\begin{equation}
F_L(k,\mu)\equiv
\sin[(L{+}1)k]-\mu^2\sin[(L{-}1)k]-2\mu\cosh(\kappa L)\sin k=0
\label{eq:EM4_secular}
\end{equation}
[Eq.~\eqref{eq:EM_SF_secular}] are real and nondegenerate, with right eigenmodes
$\langle x|\psi^{R}_{n,\ell}(\mu)\rangle\propto e^{\eta_\ell\kappa(x-1)}
[\sin((L{+}1{-}x)k_n)+\mu e^{-\eta_\ell\kappa L}\sin((x{-}1)k_n)]$
[Eq.~\eqref{eq:EM_SF_reference_modes}], the left modes being their biorthogonal duals.
Increasing $\mu$ makes two real roots of Eq.~\eqref{eq:EM4_secular} coalesce; for
generic $L$ the isolated-chain threshold follows from the double-root conditions
$F_L(k_c,\mu_c^{(0)})=\partial_kF_L(k_c,\mu_c^{(0)})=0$, which involve no strong-$\mu$
approximation, while for $L=4\ell$ the first coalescence occurs at the solvable point
$k_c=\pi/2$, giving Eq.~\eqref{eq:SF_reference_EP}. Feeding the finite-$\mu$ energies
and the projected inter-chain elements
$v_{mn}(\mu)=\langle\psi_{m,b}^{L}(\mu)|\mathcal V_{\perp}|\psi_{n,a}^{R}(\mu)\rangle
=\langle\psi_{m,a}^{L}(\mu)|\mathcal V_{\perp}|\psi_{n,b}^{R}(\mu)\rangle$
into the \emph{same} parity blocks Eq.~\eqref{eq:HN_chain_parity_blocks} used under OBCs
reproduces Eq.~\eqref{eq:SF_pair_threshold} [see also SM~\cite{suppmat}~S4]. The
boundary closure therefore leaves the activation mechanism untouched: it only deforms
the reference data on which the very same EP condition acts.

\paragraph{Reference-EP-induced threshold collapse.---}
The finite-$\mu$ example above is a particular instance of a general phenomenon that
arises whenever the \emph{reference} operator itself is tuned toward an EP. Let
$\mathcal H_0(g)$ depend on a control parameter $g$ such that a pair $(p,q)$ of its own
eigenvalues coalesces at $g=g_c$, and let $\delta g=g_c-g\to0^+$. Generic square-root
branching then fixes both ingredients of Eq.~\eqref{eq:act_ratio} simultaneously, with
opposite exponents: the levels split as $\Delta_{\rm act}=c\,\delta g^{1/2}$, while the
two eigenvectors become parallel, so the biorthogonal self-overlaps normalizing
$P_{p,q}$ vanish and the projected couplings diverge with the inverse Petermann-type
factor, $G_{\rm act}=\gamma\,\delta g^{-1/2}$ ($c,\gamma>0$), unless a symmetry enforces
a cancellation. Their ratio collapses \emph{linearly},
\begin{equation}
\lambda_c^{(p,q)}
=\frac{\Delta_{\rm act}}{G_{\rm act}}
=\frac{c}{\gamma}\left(g_c-g\right)+O\!\left[(g_c-g)^{3/2}\right],
\label{eq:EM4_linear}
\end{equation}
with a unit exponent fixed by the square-root structure alone, independently of
microscopic detail. Two consequences follow. First, only this \emph{terminal} pair is
singular, every other candidate having analytic $\Delta E$ and $G_{\rm act}$ at $g_c$;
the lower envelope Eq.~\eqref{eq:threshold_minimization} is therefore
\emph{necessarily} taken over by the terminal channel sufficiently close to $g_c$,
whatever controls the transition far from it. Second, nonmonotonic thresholds arise
naturally when a regular channel hardens with $g$ while the terminal channel softens,
in which case the trend reversal signals a change of the microscopic pair responsible
for the first instability, rather than any phase transition or nonanalyticity of
$\mathcal H(\lambda)$. The $L=10$ curve of Fig.~\ref{fig:cNHSE}(b2), with $g=\mu$ and
$g_c=\mu_c^{(0)}$, realizes precisely this through the sequence
$(1,2)\to(2,3)\to(4,5)$ [SM~\cite{suppmat}~S4]. A reference EP thus acts not as an
object to be observed, but as a resource: it amplifies the projected activation strength
and squeezes the threshold linearly to zero.

\clearpage
\newpage
\onecolumngrid
\subsection*{\normalsize Supplementary Materials for ``Exceptional activated mode theory for generalized real-complex transitions''}

\setcounter{equation}{0}
\setcounter{figure}{0}
\setcounter{table}{0}
\setcounter{section}{0}
\setcounter{page}{1}
\renewcommand{\theequation}{S\arabic{equation}}
\renewcommand{\thefigure}{S\arabic{figure}}
\renewcommand{\thesection}{S\arabic{section}}
\renewcommand{\thepage}{S\arabic{page}}

This Supplemental Material contains the following material:
\begin{enumerate}
\item A full-spectrum test showing that the activated-mode threshold is not a finite-order weak-coupling expansion.
\item A discussion of the two-mode projection and the neglected admixture with remote reference modes.
\item A derivation of the selected $(1,2)$ activated channel in the coupled Hatano--Nelson (HN) ladder, including its reflection-related partner $(L-1,L)$.
\item A finite-$\mu$ analysis of the impurity-closed coupled HN ladder, including the exact reference spectrum, the isolated-chain threshold $\mu_c^{(0)}$, and the boundary-induced activated-channel crossovers responsible for the nonmonotonic $\lambda_c(\mu)$. 
\item A derivation of the topological-edge-channel threshold in the locally activated SSH chain.
\end{enumerate}

\section{Benchmarking the activated-mode threshold in the strong-coupling regime}

The main text introduced a two-mode active block [Eq.~(2)] and obtained each pair-resolved threshold from its exact EP discriminant [Eqs.~(3)--(5)], without expanding the EP condition in $\lambda$. It then applied this construction to the coupled HN ladder of Eqs.~(8)--(10). The goal of this section is to test that prediction directly against the full spectrum when the transition lies outside a parametrically weak-coupling regime. For small system sizes, the first real-to-complex transition occurs at $\lambda_c=\mathcal O(10^{-1})$--$\mathcal O(1)$, comparable to the single-chain hopping scale.

For each finite $L$, we compute the spectrum of the full $2L\times2L$ Hamiltonian matrix
\begin{equation}
    \mathcal{H}_{\text{coupled-HN}}(\lambda)
    =
    \mathcal{H}_0+\lambda\mathcal{V}_{\perp}
\end{equation}
by full numerical diagonalization at each trial value of $\lambda$. The transition threshold is extracted from
\begin{equation}
f(\lambda)
=
\max_n |\operatorname{Im}E_n(\lambda)|,
\label{eq:S1_imag_indicator}
\end{equation}
as the first value of $\lambda$ at which $f(\lambda)$ exceeds a small numerical tolerance,
\begin{equation}
    \lambda_c^{\rm num}
    =
    \inf\{\lambda>0\,|\,f(\lambda)>\epsilon\}.
    \label{eq:S1_num_threshold}
\end{equation}
In the numerical results shown below, we use $\epsilon=10^{-10}$. 
Because finite systems can exhibit a re-entrant real-spectrum window at larger $\lambda$, the threshold is not obtained by assuming that the spectrum remains complex for all $\lambda>\lambda_c$. 
Instead, we first scan $\lambda$ to locate the first real-to-complex crossing and then refine this crossing by bisection.
\begin{figure}[!htp]
    \centering
    \includegraphics[width=0.4\linewidth]{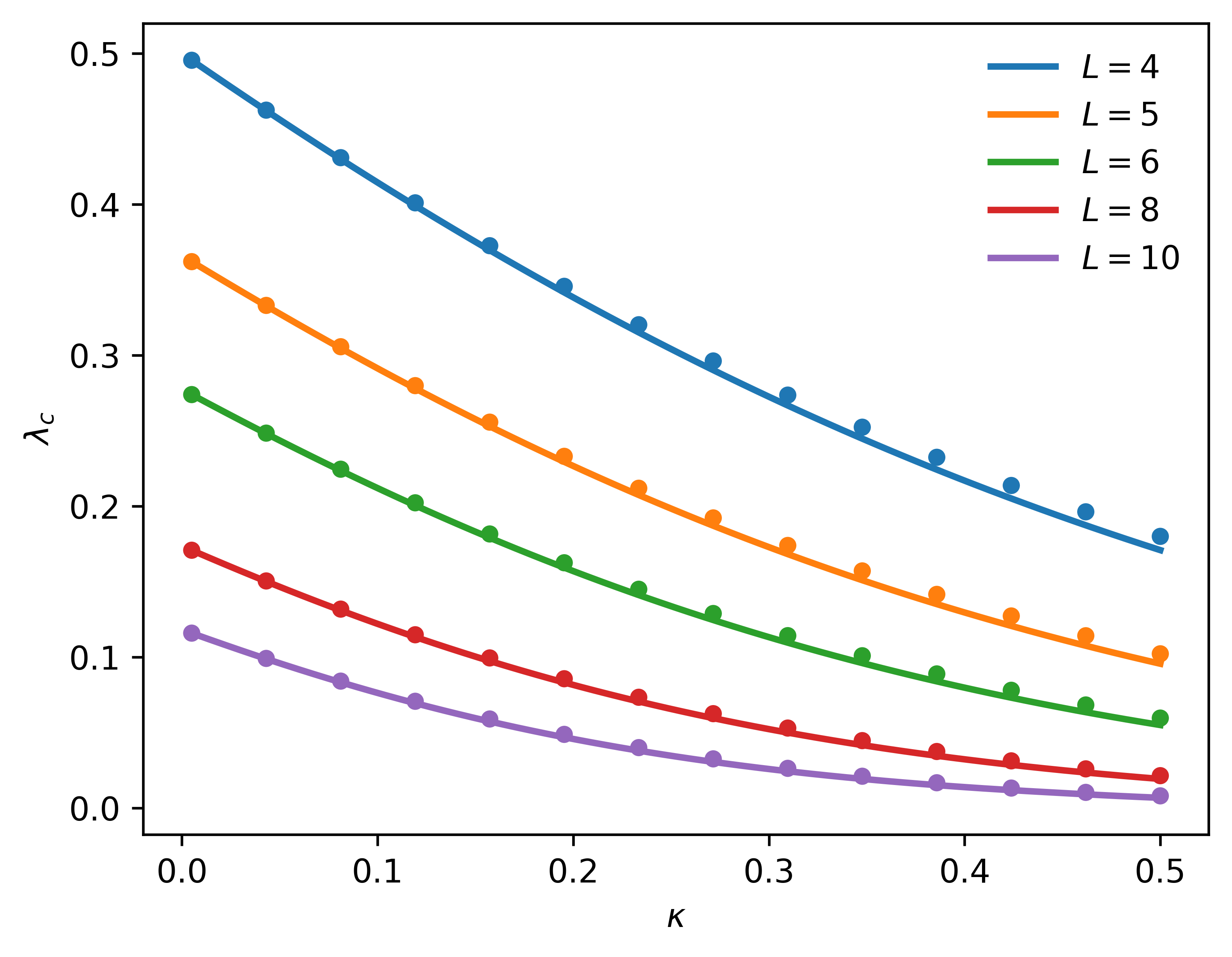}
    \caption{
    Accuracy of the two-mode prediction beyond weak coupling for the coupled HN ladder of main-text Eqs.~(8)--(10). Dots are full numerical-diagonalization thresholds $\lambda_c^{\rm num}$ extracted from Eq.~\eqref{eq:S1_num_threshold}; solid curves are the activated-mode lower-envelope prediction $\lambda_c$ from Eq.~\eqref{eq:S1_AMT_lower_envelope}. 
    Different colors correspond to different system sizes $L$, as indicated in the legend. 
    The agreement persists for $\lambda_c=\mathcal O(10^{-1})$--$\mathcal O(1)$, showing that the active-mode threshold is not restricted to an infinitesimal-coupling regime. 
    Thus the two-mode prediction remains accurate for $\lambda_c=\mathcal O(10^{-1})$--$\mathcal O(1)$, outside the parametrically weak-coupling regime. The point $\kappa=0$ is excluded because the ladder is then Hermitian for all real $\lambda$ and has no real-to-complex threshold.
    }
    \label{fig:S1_large_lambda}
\end{figure}

For the activated-mode result we evaluate the pair-resolved threshold in Eq.~\eqref{eq:S2_HN_two_branch_threshold} exactly, without expanding in $\lambda$, and minimize it over all admissible pairs. Before an isolated second-order EP, only adjacent branches in the ordered spectrum can merge. Since $E_n=2\cos[n\pi/(L+1)]$ decreases monotonically with $n$, and the inter-chain coupling splits each degenerate spatial level into two branch labels, this spectral-adjacency requirement selects the branch-changing pairs with $i+j$ odd. We therefore compute $\lambda_c^{(i,j)}$ for every $i<j$ with $i+j$ odd and take the lower envelope
\begin{equation}
    \lambda_c
    =
    \min_{\substack{i<j\\ i+j\ {\rm odd}}}
    \lambda_c^{(i,j)}.
    \label{eq:S1_AMT_lower_envelope}
\end{equation}
Thus Fig.~\ref{fig:S1_large_lambda} compares the transition threshold of the full $2L\times2L$ matrix with the active-mode lower envelope. 
No expansion in $\lambda$ is used in Eq.~\eqref{eq:S1_AMT_lower_envelope}; the coupling enters through the exact exceptional-point condition of the active block.

Figure~\ref{fig:S1_large_lambda} shows that $\lambda_c^{\rm num}$ agrees closely with $\lambda_c$ for several finite sizes. 
The most important feature is not only the small discrepancy, but the scale of the threshold: for the smallest systems, the transition occurs at order-one fractions of the hopping amplitude. 
Therefore the agreement cannot be interpreted as a weak-coupling coincidence. 
Rather, it confirms the logic used in the main text: once the relevant active pair is selected, the threshold is fixed by the exceptional-point condition inside that finite active subspace.

This also clarifies the relation between activated-mode theory and ordinary perturbative reasoning. 
The calculation does not assume that $\lambda$ is small compared with the unperturbed level spacing or with the hopping scale. 
The reference spectrum of $\mathcal{H}_0$ is used only to organize candidate active channels and to compute their projected matrix elements. 
The actual threshold is then obtained from the vanishing discriminant of the active block, which is solved exactly in $\lambda$. 
Consequently, the method remains applicable when the first transition occurs at large coupling, provided the transition is dominated by a well-defined low-dimensional active channel.

\section{Validity of the two-mode projection: neglected admixture with remote reference modes}
\label{sec:S_validity}

The main text projected the full problem onto a candidate pair and solved the resulting $2\times2$ active block exactly in $\lambda$ [Eqs.~(2)--(5)]. That construction retains the selected pair but omits its virtual admixture with all other reference modes. The goal of this section is to define those remote modes and state a quantitative criterion for when the two-mode projection is accurate. For a selected active pair $(p,q)$, we call all reference modes $r\notin\{p,q\}$ \emph{remote reference modes}. With biorthogonal eigenmodes $\langle\psi_i^L|\psi_j^R\rangle=\delta_{ij}$ and projected elements $V_{ij}=\langle\psi_i^L|\mathcal V|\psi_j^R\rangle$, the direct active block is
\begin{equation}
\mathcal H_{pq}^{\rm act}(\lambda)=
\begin{pmatrix}
E_p+\lambda D_p & \lambda A_{pq}\\
\lambda B_{qp} & E_q+\lambda D_q
\end{pmatrix},
\quad
D_i=V_{ii},\quad A_{pq}=V_{pq},\quad B_{qp}=V_{qp}.
\label{eq:Sx_active_block}
\end{equation}
Equation~\eqref{eq:Sx_active_block} is exactly $P\mathcal H(\lambda)P$ with $P=P_p+P_q$. It is not the exact restriction of the full spectrum to that pair, because $\mathcal V$ generically couples $(p,q)$ to the remote modes. The effective Hamiltonian for the branches connected to $(p,q)$ as $\lambda\to0$ instead follows from a Löwdin/Feshbach downfolding,
\begin{equation}
\begin{aligned}
\mathcal{H}_{\rm eff}(E,\lambda)
={}&P\mathcal{H}(\lambda)P
+\lambda^2P\mathcal{V}Q
\left(E-Q\mathcal{H}_0Q\right)^{-1}Q\mathcal{V}P\\
&+O(\lambda^3),
\qquad Q=\mathbf{1}-P,
\end{aligned}
    \label{eq:Sx_downfold}
\end{equation}
where the second term is the leading virtual-admixture correction absent from Eq.~\eqref{eq:Sx_active_block}. Evaluating the resolvent near the active EP energy $E_\ast$ shifts its matrix elements by
\begin{equation}
    \delta\mathcal{H}_{ij}
    =
    \lambda^2\sum_{r\neq p,q}
    \frac{A_{ir}B_{rj}}{E_\ast-E_r},
    \qquad
    i,j\in\{p,q\},
    \label{eq:Sx_correction}
\end{equation}
with $A_{ir}=\langle\psi_i^L|\mathcal{V}|\psi_r^R\rangle$ and $B_{rj}=\langle\psi_r^L|\mathcal{V}|\psi_j^R\rangle$. This second-order admixture renormalizes both the diagonal shifts and the off-diagonal elements $A_{pq},B_{qp}$.

The size of Eq.~\eqref{eq:Sx_correction} is controlled not only by the explicit factor $\lambda^2$, but also by spectral isolation and the projected matrix elements. Define the retained active scale at the EP by
\begin{equation}
s_{pq}\equiv
\max\!\left\{
|\Delta E_{pq}|,
|\lambda_c\Delta D_{pq}|,
\lambda_c\sqrt{-A_{pq}B_{qp}}
\right\},
\qquad \Delta D_{pq}=D_p-D_q,
\label{eq:Sx_active_scale}
\end{equation}
whose three entries have a common parametric order when the terms in the discriminant of Eq.~\eqref{eq:Sx_active_block} balance. The two-mode reduction is reliable when
\begin{equation}
    \|\delta\mathcal{H}(E_\ast,\lambda_c)\|
    \ll
    s_{pq},
    \label{eq:Sx_smallness_condition}
\end{equation}
Explicitly, the energy-dependent correction is
\begin{equation}
    \delta\mathcal{H}_{ij}(E_\ast,\lambda_c)
    =
    \lambda_c^2
    \sum_{r\notin\{p,q\}}
    \frac{
    \langle\psi_i^L|\mathcal{V}|\psi_r^R\rangle
    \langle\psi_r^L|\mathcal{V}|\psi_j^R\rangle
    }{E_\ast-E_r},
    \qquad
    i,j\in\{p,q\}.
    \label{eq:Sx_correction_general}
\end{equation}
For spectrally isolated remote modes with bounded projected couplings, Eq.~\eqref{eq:Sx_correction_general} is of order $\lambda_c^2$ times an $\mathcal O(1)$ coefficient, while the retained active-block terms are of order $\lambda_c\sqrt{-A_{pq}B_{qp}}$ or $|\Delta E_{pq}|$. In that common case the fractional correction scales as
\begin{equation}
    \frac{\|\delta\mathcal{H}\|}{s_{pq}}
    \sim
    \lambda_c
    \frac{C_{pq}}{\sqrt{-A_{pq}B_{qp}}},
    \label{eq:Sx_fractional}
\end{equation}
where the remote-mode coefficient is defined explicitly as
\begin{equation}
C_{pq}\equiv
\max_{i,j\in\{p,q\}}
\left|
\sum_{r\notin\{p,q\}}
\frac{A_{ir}B_{rj}}{E_\ast-E_r}
\right|.
\label{eq:Sx_Cpq}
\end{equation}
For example, if every remote level obeys $|E_\ast-E_r|\geq\Delta_{\rm rem}>0$, the projected couplings are bounded, and their finite sum does not grow with the scaling limit, then $C_{pq}=\mathcal O(1)$. The explicit validity criterion is therefore
\begin{equation}
\boxed{
\lambda_c\,
\frac{C_{pq}}{\sqrt{-A_{pq}B_{qp}}}
\ll1 .}
\label{eq:Sx_validity_criterion}
\end{equation}
Whenever this fractional correction is small at the EP, the two-mode reduction is quantitatively accurate. It is automatically parametrically small when $\lambda_c$ itself is small and $C_{pq}/\sqrt{-A_{pq}B_{qp}}$ remains bounded, as in both cNHSE regimes considered here: the large-$\kappa L$ threshold is exponentially small and the small-$\kappa L$ finite-size threshold is algebraically small.

The large-coupling window of Sec.~S1, $\lambda_c=\mathcal O(10^{-1})$--$\mathcal O(1)$, is not covered by this small-$\lambda_c$ scaling argument. There the two-mode approximation is instead validated a posteriori by Fig.~\ref{fig:S1_large_lambda}, which compares the lower-envelope prediction directly with full numerical diagonalization of the original $2L\times2L$ matrix. In summary, Sec.~S1 establishes numerical accuracy beyond weak coupling, whereas this section explains analytically why the reduction is controlled whenever Eq.~\eqref{eq:Sx_validity_criterion} is small.

\section{Selection of the $(1,2)$ activated channel in the coupled Hatano--Nelson ladder}
\label{sec:S3_HN_selection}

The main text applied the general pair-resolved EP condition to the OBC coupled HN ladder [Eqs.~(8)--(10)] and reported the closed selected-channel threshold in Eq.~(11). This section supplies the mode projection and minimization that lead to that result. In particular, we show that the first real-to-complex transition is controlled by the upper band-edge pair $(p,q)=(1,2)$, together with its spectrally reflected partner $(L-1,L)$.

We consider two oppositely nonreciprocal open chains with a local inter-chain coupling,
\begin{equation}
    \mathcal{H}(\lambda)
    =
    \mathcal{H}_{0}
    +
    \lambda \mathcal{V}_{\perp},
    \label{eq:S2_HN_model}
\end{equation}
with
\begin{equation}
\begin{aligned}
    \mathcal{H}_{0}
    =
    \sum_{x=1}^{L-1}
    \Big[
    e^{\kappa}a_{x+1}^{\dagger}a_x^{\vphantom{\dagger}}
    +
    e^{-\kappa}a_x^{\dagger}a_{x+1}^{\vphantom{\dagger}}
    +
    e^{-\kappa}b_{x+1}^{\dagger}b_x^{\vphantom{\dagger}}
    +
    e^{\kappa}b_x^{\dagger}b_{x+1}^{\vphantom{\dagger}}
    \Big],
\end{aligned}
    \label{eq:S2_HN_H0}
\end{equation}
and
\begin{equation}
    \mathcal{V}_{\perp}
    =
    \sum_{x=1}^{L}
    \left(
    a_x^{\dagger}b_x^{\vphantom{\dagger}}
    +
    b_x^{\dagger}a_x^{\vphantom{\dagger}}
    \right).
    \label{eq:S2_HN_Vperp}
\end{equation}
The two chains have opposite inverse skin lengths $\pm\kappa$. Each chain can be mapped by a non-unitary gauge transformation to the same reciprocal open chain. Therefore the reference spatial eigenmodes are
\begin{equation}
    \phi_j(x)
    =
    \sqrt{\frac{2}{L+1}}
    \sin\frac{j\pi x}{L+1},
    \qquad
    j=1,\ldots,L,
    \label{eq:S2_phi_j}
\end{equation}
with real energies
\begin{equation}
    E_j
    =
    2\cos\frac{j\pi}{L+1}.
    \label{eq:S2_E_j}
\end{equation}
Before the inter-chain coupling is applied, each spatial level $j$ is twofold degenerate, with one copy on each chain. Projecting $\mathcal V_\perp$ onto these two copies gives the intra-mode block $E_j\mathbbm 1_2+\lambda\left(\begin{smallmatrix}0&v_{jj}\\v_{jj}&0\end{smallmatrix}\right)$, whose symmetric and antisymmetric chain combinations split into two real branches. Retaining spatial modes $i$ and $j$ gives the same four-mode construction and $2\times2$ parity blocks displayed explicitly in Eqs.~\eqref{eq:S3_four_mode_block}--\eqref{eq:S3_reduced_blocks} below. The first upper-edge instability is a collision between the lower branch of mode $i$ and the upper branch of mode $j$. In this branch convention its active block is
\begin{equation}
    \mathcal{H}_{ij}^{\rm act}(\lambda)
    =
    \begin{pmatrix}
    E_i+\lambda D_i & \lambda A_{ij}\\
    \lambda B_{ji} & E_j+\lambda D_j
    \end{pmatrix},
    \label{eq:S2_active_block}
\end{equation}
where
\begin{equation}
    D_i\equiv\langle\psi_i^L|\mathcal V_\perp|\psi_i^R\rangle,
    \qquad
    \Delta D_{ij}=D_i-D_j.
    \label{eq:S2_D_definition}
\end{equation}
The exceptional-point condition is the vanishing of the discriminant,
\begin{equation}
    \left[
    \Delta E_{ij}
    +
    \lambda \Delta D_{ij}
    \right]^2
    +
    4\lambda^2 A_{ij}B_{ji}
    =
    0,
    \label{eq:S2_EP_condition}
\end{equation}
with $\Delta E_{ij}=E_i-E_j$. When the reduced matrix elements are real and $A_{ij}B_{ji}<0$, Eq.~\eqref{eq:S2_EP_condition} gives two candidate EP branches,
\begin{equation}
    \lambda_{c,\sigma}^{(i,j)}
    =
    -\frac{\Delta E_{ij}}
    {\Delta D_{ij}-2\sigma\sqrt{-A_{ij}B_{ji}}},
    \qquad
    \sigma=\pm .
    \label{eq:S2_general_branch}
\end{equation}

We now evaluate the quantities entering Eq.~\eqref{eq:S2_general_branch}. Direct projection of the local inter-chain coupling gives
\begin{equation}
\begin{aligned}
    \Delta D_{ij}
    &=
    -\frac{2e^{\kappa(L+1)}}{L+1}
    \sum_{x=1}^{L}
    e^{-2\kappa x}
    \left[
    \sin^2\frac{i\pi x}{L+1}
    +
    \sin^2\frac{j\pi x}{L+1}
    \right],
    \\
    A_{ij}
    &=
    \frac{2e^{\kappa(L+1)}}{L+1}
    \sum_{x=1}^{L}
    e^{-2\kappa x}
    \sin\frac{i\pi x}{L+1}
    \sin\frac{j\pi x}{L+1},
    \\
    B_{ji}
    &=
    -A_{ij},
    \\
    \Delta E_{ij}
    &=
    2\cos\frac{i\pi}{L+1}
    -
    2\cos\frac{j\pi}{L+1}.
\end{aligned}
\label{eq:S2_HN_projected_elements}
\end{equation}
Substituting Eq.~\eqref{eq:S2_HN_projected_elements} into Eq.~\eqref{eq:S2_general_branch} gives the pair-resolved threshold
\begin{equation}
\begin{aligned}
    \lambda_{c,\sigma}^{(i,j)}
    &=(L+1)e^{-\kappa(L+1)}
    \frac{\displaystyle
    \cos\frac{i\pi}{L+1}-\cos\frac{j\pi}{L+1}}
    {W_{ij}^{(\sigma)}},\\
    W_{ij}^{(\sigma)}
    &=\sum_{x=1}^{L}e^{-2\kappa x}
    \left[
    \sin^2\frac{i\pi x}{L+1}
    +\sin^2\frac{j\pi x}{L+1}
    \right]\\
    &\quad+2\sigma\left|
    \sum_{x=1}^{L}e^{-2\kappa x}
    \sin\frac{i\pi x}{L+1}
    \sin\frac{j\pi x}{L+1}
    \right|,
    \qquad \sigma=\pm.
\end{aligned}
\label{eq:S2_HN_two_branch_threshold}
\end{equation}
For the upper band-edge ordering $i<j$, the physical positive threshold is obtained from the smaller positive branch. In the parameter regime considered here this is the $\sigma=+$ branch. Therefore the active-channel selection is obtained by minimizing Eq.~\eqref{eq:S2_HN_two_branch_threshold} over candidate pairs,
\begin{equation}
    \lambda_c
    =
    \min_{i<j,\;i+j\,{\rm odd}}
    \lambda_c^{(i,j)}.
    \label{eq:S2_lower_envelope}
\end{equation}
The parity condition selects the branch-changing channels relevant to the first upper-edge EP collision.

It remains to show why the minimum is attained by $(p,q)=(1,2)$. Define
\begin{equation}
    \theta=\frac{\pi}{L+1},
    \qquad
    r=e^{-2\kappa},
    \qquad
    s_j(x)=\sin(j\theta x).
\end{equation}
For $i<j$, Eq.~\eqref{eq:S2_HN_two_branch_threshold} can be written as
\begin{equation}
    \lambda_c^{(i,j)}
    =
    (L+1)e^{-\kappa(L+1)}
    \frac{\delta_{ij}}{W_{ij}},
    \label{eq:S2_pair_threshold_compact}
\end{equation}
where
\begin{equation}
    \delta_{ij}
    =
    \cos(i\theta)-\cos(j\theta)>0,
    \label{eq:S2_delta_def}
\end{equation}
and
\begin{equation}
\begin{aligned}
    W_{ij}
    &=
    \sum_{x=1}^{L}
    r^x
    \left[
    s_i^2(x)+s_j^2(x)
    \right]
    \\
    &\quad+
    2
    \left|
    \sum_{x=1}^{L}
    r^x s_i(x)s_j(x)
    \right| .
\end{aligned}
\label{eq:S2_W_def}
\end{equation}
The common prefactor $(L+1)e^{-\kappa(L+1)}$ in Eq.~\eqref{eq:S2_pair_threshold_compact} does not affect the minimizer. Determining which modes are activated is therefore reduced to minimizing
\begin{equation}
    F_{ij}=\frac{\delta_{ij}}{W_{ij}}.
    \label{eq:S2_Fpq}
\end{equation}

First, there is an exact reflection symmetry. Using
\begin{equation}
    s_{L+1-j}(x)
    =
    \sin[(L+1-j)\theta x]
    =
    (-1)^{x+1}s_j(x),
\end{equation}
one obtains
\begin{equation}
    W_{L+1-j,L+1-i}=W_{ij}.
\end{equation}
At the same time,
\begin{equation}
\begin{aligned}
    \delta_{L+1-j,L+1-i}
    &=
    \cos[(L+1-j)\theta]
    -
    \cos[(L+1-i)\theta]
    \\
    &=
    -\cos(j\theta)+\cos(i\theta)
    =
    \delta_{ij}.
\end{aligned}
\end{equation}
Therefore
\begin{equation}
    \lambda_c^{(L+1-j,L+1-i)}
    =
    \lambda_c^{(i,j)} .
    \label{eq:S2_reflection_symmetry}
\end{equation}

Second, the detuning strongly favors adjacent band-edge pairs. The identity
\begin{equation}
    \delta_{ij}
    =
    2\sin\frac{(i+j)\theta}{2}
    \sin\frac{(j-i)\theta}{2}
    \label{eq:S2_detuning_identity}
\end{equation}
shows that the smallest detuning comes from adjacent pairs $j=i+1$, and among them from the band edges. For adjacent pairs,
\begin{equation}
    \delta_{j,j+1}
    =
    2\sin\frac{(2j+1)\theta}{2}
    \sin\frac{\theta}{2},
    \label{eq:S2_adjacent_detuning}
\end{equation}
which is smallest at $j=1$ and, by reflection, at $j=L-1$.

The only possible way for another pair to win would be for its projected coupling $W_{ij}$ to overcome this larger detuning. This does not occur in the finite-size crossover regime of interest. To see the scaling explicitly, take
\begin{equation}
    \chi=\kappa(L+1)
\end{equation}
fixed as $L$ increases. For fixed low-lying indices $i,j$,
\begin{equation}
    W_{ij}
    =
    (L+1)\mathcal{I}_{ij}(\chi)+O(1),
    \label{eq:S2_W_scaling}
\end{equation}
with
\begin{equation}
\begin{aligned}
    \mathcal{I}_{ij}(\chi)
    =
    \int_0^1
    e^{-2\chi u}
    \Big[
    &\sin^2(i\pi u)+\sin^2(j\pi u)
    \\
    &+
    2|\sin(i\pi u)\sin(j\pi u)|
    \Big]du .
\end{aligned}
\label{eq:S2_Ipq}
\end{equation}
The detuning scales as
\begin{equation}
    \delta_{ij}
    =
    \frac{\pi^2(j^2-i^2)}{2(L+1)^2}
    +
    O(L^{-4}).
    \label{eq:S2_delta_scaling}
\end{equation}
Hence
\begin{equation}
    \lambda_c^{(i,j)}
    =
    \frac{\pi^2e^{-\chi}}{2(L+1)^2}
    \frac{j^2-i^2}{\mathcal{I}_{ij}(\chi)}
    +
    O(L^{-3}).
    \label{eq:S2_scaling_threshold}
\end{equation}
This expression makes the selection mechanism transparent. The band-edge pair $(1,2)$ has the smallest possible detuning factor, $j^2-i^2=3$, while the projected coupling changes only by an order-one factor among nearby low-lying pairs. Thus the lower envelope selects $(1,2)$, with the reflected partner $(L-1,L)$ fixed by Eq.~\eqref{eq:S2_edge_degeneracy}. Direct finite-sum minimization of Eq.~\eqref{eq:S2_HN_two_branch_threshold} confirms this selection for the parameter regime used in the numerical comparison.

Finally, substituting $(p,q)=(1,2)$ into Eq.~\eqref{eq:S2_HN_two_branch_threshold} gives a closed-form selected-channel threshold. Define
\begin{equation}
    C_m
    =
    \sum_{x=1}^{L}
    e^{-2\kappa x}
    \cos\frac{m\pi x}{L+1}.
    \label{eq:S2_Cm_def}
\end{equation}
Then
\begin{equation}
    \lambda_c
    =
    \frac{
    4(L+1)e^{-\kappa(L+1)}
    \sin\frac{3\pi}{2(L+1)}
    \sin\frac{\pi}{2(L+1)}
    }
    {
    2C_0+2C_1-C_2-2C_3-C_4
    }.
    \label{eq:S2_HN_closed_threshold}
\end{equation}
This is the finite-size active-channel threshold for the coupled HN ladder.

Two limiting forms are useful. For $\kappa L\gg1$,
\begin{equation}
    \lambda_c
    \simeq
    \frac{2}{3}
    (L+1)e^{-\kappa(L+1)}
    \frac{\sinh^3\kappa}{\cosh\kappa},
    \qquad
    \kappa L\gg1 .
    \label{eq:S2_HN_large_kL}
\end{equation}
This limit, which corresponds to an exponentially small coupling threshold, is governed by NHSE feedback tunneling through the inter-chain coupling, as in the conventional cNHSE. In the opposite finite-size or weak-NHSE regime, $\kappa L\ll1$ with nonzero $\kappa$, the $L$ dependence is dictated primarily by the energetic spacings and crosses over to the power law
\begin{equation}
    \lambda_c
    \simeq
    \frac{3\pi^2}{2(L+1)^2},
    \qquad
    \kappa L\ll1 .
    \label{eq:S2_HN_small_kL}
\end{equation}
The point $\kappa=0$ itself is singular for this real-to-complex problem because the ladder is Hermitian for real $\lambda$. Equation~\eqref{eq:S2_HN_small_kL} is the leading finite-size crossover at small but nonzero nonreciprocity; its leading coefficient is $\kappa$ independent, while $\kappa$-dependent corrections enter at higher order in $\kappa L$. Equations~\eqref{eq:S2_HN_closed_threshold}--\eqref{eq:S2_HN_small_kL} show that the same activated pair controls both the conventional exponentially small cNHSE threshold and the algebraic finite-size band-edge threshold through the unified expression~\eqref{eq:S2_HN_closed_threshold}, even though the two limits arise from different physical mechanisms.

\section{Finite-$\mu$ impurity-closed active-channel crossover}
\label{sec:S4_SF_selection}

The main text closed the HN ladder by end-to-end impurity bonds [Eq.~(14)], evaluated the same pair-resolved EP condition using the finite-$\mu$ reference modes, and obtained the lower-envelope threshold in Eq.~(19). It found that varying $\mu$ can switch the activated pair and produce a nonmonotonic threshold before the reference chain itself becomes complex. This section gives the complete finite-$\mu$ derivation. The parameter $\mu$ is the strength of the end-to-end impurity bond: $\mu=0$ gives OBCs and $\mu=1$ gives the nominal periodic bond strength. We use ``impurity-PBC" for this tunable closure, but $\mu\neq1$ is not a strictly periodic boundary condition.

Two distinct thresholds must be separated. The quantity $\mu_{c}^{(0)}$ denotes the first real-to-complex transition of an isolated impurity-closed chain at $\lambda=0$. By contrast, for a fixed
\begin{equation}
0\leq\mu<\mu_{c}^{(0)},
\label{eq:S3_mu_regime}
\end{equation}
the reference spectrum remains real, and we ask for the first inter-chain coupling $\lambda_c(\mu)$ that makes the coupled ladder complex. The boundary closure deforms the reference energies and eigenvectors continuously from their open-chain values. The nondegenerate OBC spectrum at $\mu=0$ therefore remains isolated under sufficiently small $\mu$. Consequently, an infinitesimal impurity-PBC hopping does not generically cause an immediate jump of the activated channel; a switch occurs only when two pair-resolved thresholds cross on the global lower envelope.

\subsection{Impurity-closed coupled Hatano--Nelson ladder}

We write the coupled ladder as
\begin{equation}
\mathcal H_{\rm imp}(\lambda,\mu)=
\begin{pmatrix}
H_b(\mu)&\lambda\mathbbm 1_L\\
\lambda\mathbbm 1_L&H_a(\mu)
\end{pmatrix}
=\mathcal H_{0}^{\rm imp}(\mu)+\lambda\mathcal V_{\perp},
\label{eq:S3_full_model}
\end{equation}
where the two oppositely nonreciprocal chains are
\begin{equation}
\begin{aligned}
H_b(\mu)={}&\sum_{x=1}^{L-1}\left(e^{\kappa}|x\rangle\langle x+1|+e^{-\kappa}|x+1\rangle\langle x|\right)+\mu\left(e^{-\kappa}|1\rangle\langle L|+e^{\kappa}|L\rangle\langle1|\right),
\\
H_a(\mu)={}&\sum_{x=1}^{L-1}\left(e^{-\kappa}|x\rangle\langle x+1|+e^{\kappa}|x+1\rangle\langle x|\right)+\mu\left(e^{\kappa}|1\rangle\langle L|+e^{-\kappa}|L\rangle\langle1|\right).
\end{aligned}
\label{eq:S3_single_chain_model}
\end{equation}
The local inter-chain coupling is
\begin{equation}
\mathcal V_{\perp}=\sum_{x=1}^{L}\left(a_x^\dagger b_x+b_x^\dagger a_x\right),
\label{eq:S3_Vperp}
\end{equation}
and connects equal spatial sites on the two chains. Equations~\eqref{eq:S3_full_model}--\eqref{eq:S3_Vperp} reproduce the complete impurity-closed model locally within the Supplemental Material.

At $\mu=0$, the two chains are mapped by opposite non-unitary gauge transformations to the same reciprocal open chain and therefore have identical real reference energies. For every finite $\mu$, the spectra of chains $a$ and $b$ remain equal to each other because they are related by the mirror operation $x\mapsto L+1-x$ together with $\kappa\mapsto-\kappa$. This does not mean that either spectrum is independent of $\mu$: the end-to-end bonds continuously deform both their common energies and their biorthogonal eigenmodes from the open-chain values.

\subsection{Exact finite-$\mu$ impurity-chain spectrum}

Consider a right eigenstate of the $b$ chain,
\begin{equation}
H_b(\mu)|\psi_n^R\rangle
=
E_n|\psi_n^R\rangle.
\end{equation}
For the bulk sites $x=2,\ldots,L-1$, its site amplitudes satisfy
\begin{equation}
e^{\kappa}\psi_n^R(x+1)
+
e^{-\kappa}\psi_n^R(x-1)
=
E_n\psi_n^R(x).
\end{equation}
Writing
\begin{equation}
\psi_n^R(x)
=
e^{-\kappa(x-1)}\phi_n(x)
\end{equation}
removes the nonreciprocity from the bulk recurrence and gives
\begin{equation}
\phi_n(x+1)+\phi_n(x-1)
=
E_n\phi_n(x).
\label{eq:S3_bulk_recurrence}
\end{equation}
For a solution with momentum $k_n$, Eq.~\eqref{eq:S3_bulk_recurrence} gives
\begin{equation}
E_n=2\cos k_n.
\label{eq:S3_energy_dispersion}
\end{equation}
The eigenvalue equations at the two boundary sites are
\begin{equation}
\begin{aligned}
E_n\psi_n^R(1)
&=
e^{\kappa}\psi_n^R(2)
+
\mu e^{-\kappa}\psi_n^R(L),
\\
E_n\psi_n^R(L)
&=
e^{-\kappa}\psi_n^R(L-1)
+
\mu e^{\kappa}\psi_n^R(1).
\end{aligned}
\end{equation}
After the gauge transformation, these become
\begin{equation}
\begin{aligned}
E_n\phi_n(1)
&=
\phi_n(2)
+
\mu e^{-\kappa L}\phi_n(L),
\\
E_n\phi_n(L)
&=
\phi_n(L-1)
+
\mu e^{\kappa L}\phi_n(1).
\end{aligned}
\label{eq:S3_transformed_boundary_equations}
\end{equation}
The general solution of the reciprocal bulk recurrence can be expressed in the sine basis as
\begin{equation}
\phi_n(x)
=
A_n\sin[(L+1-x)k_n]
+
B_n\sin[(x-1)k_n].
\label{eq:S3_general_bulk_solution}
\end{equation}
Both terms independently satisfy Eq.~\eqref{eq:S3_bulk_recurrence}. In particular,
\begin{equation}
\begin{aligned}
\phi_n(1)
&=
A_n\sin(Lk_n),
\\
\phi_n(L)
&=
A_n\sin k_n
+
B_n\sin[(L-1)k_n],
\\
\phi_n(L-1)
&=
A_n\sin(2k_n)
+
B_n\sin[(L-2)k_n].
\end{aligned}
\label{eq:S3_general_solution_boundaries}
\end{equation}
Substituting Eq.~\eqref{eq:S3_general_solution_boundaries} into the second boundary equation in Eq.~\eqref{eq:S3_transformed_boundary_equations} gives
\begin{equation}
2\cos k_n
\Big[
A_n\sin k_n
+
B_n\sin[(L-1)k_n]
\Big]
=
A_n\sin(2k_n)
+
B_n\sin[(L-2)k_n]
+
\mu e^{\kappa L}A_n\sin(Lk_n).
\end{equation}
Using
\begin{equation}
2\cos k_n\sin k_n
=
\sin(2k_n)
\end{equation}
and
\begin{equation}
\begin{aligned}
2\cos k_n\sin[(L-1)k_n]
=
\sin(Lk_n)+
\sin[(L-2)k_n],
\end{aligned}
\end{equation}
the boundary equation reduces to
\begin{equation}
B_n\sin(Lk_n)
=
\mu e^{\kappa L}A_n\sin(Lk_n).
\end{equation}
For a generic root, this fixes the relative coefficient to be
\begin{equation}
B_n
=
\mu e^{\kappa L}A_n.
\label{eq:S3_coefficient_relation}
\end{equation}
The isolated cases with $\sin(Lk_n)=0$ are obtained from the same expression by continuity. Absorbing the overall coefficient $A_n$ into a nonzero normalization factor $\mathcal N_n$, the transformed eigenmode can therefore be written as
\begin{equation}
\phi_n(x)
=
\mathcal N_n
\left[
\sin[(L+1-x)k_n]
+
\mu e^{\kappa L}\sin[(x-1)k_n]
\right]
\label{eq:S3_transformed_mode}
\end{equation}
which is a superposition of traveling waves in both directions, with one direction weighted by the asymmetric propagation factor $\mu e^{\kappa L}$.
This form automatically satisfies the second boundary equation. To impose the remaining boundary equation, we use
\begin{equation}
\begin{aligned}
\phi_n(1)
&=
\mathcal N_n\sin(Lk_n),
\\
\phi_n(2)
&=
\mathcal N_n
\left[
\sin[(L-1)k_n]
+
\mu e^{\kappa L}\sin k_n
\right],
\\
\phi_n(L)
&=
\mathcal N_n
\left[
\sin k_n
+
\mu e^{\kappa L}\sin[(L-1)k_n]
\right].
\end{aligned}
\end{equation}
Substituting these expressions into the first boundary equation in Eq.~\eqref{eq:S3_transformed_boundary_equations} yields
\begin{equation}
\begin{aligned}
2\cos k_n\sin(Lk_n)
={}&
\sin[(L-1)k_n]
+
\mu e^{\kappa L}\sin k_n
+
\mu e^{-\kappa L}
\left[
\sin k_n
+
\mu e^{\kappa L}\sin[(L-1)k_n]
\right]
\\
={}&
(1+\mu^2)\sin[(L-1)k_n]
+
\mu
\left(
e^{\kappa L}+e^{-\kappa L}
\right)
\sin k_n.
\end{aligned}
\end{equation}
Finally, using
\begin{equation}
2\cos k_n\sin(Lk_n)
=
\sin[(L+1)k_n]
+
\sin[(L-1)k_n]
\end{equation}
and
\begin{equation}
e^{\kappa L}+e^{-\kappa L}
=
2\cosh(\kappa L),
\end{equation}
we obtain the exact finite-size quantization condition
\begin{equation}
\begin{aligned}
F_L(k,\mu)
\equiv{}&
\sin[(L+1)k]
-
\mu^2\sin[(L-1)k]
-
2\mu\cosh(\kappa L)\sin k
=
0.
\end{aligned}
\label{eq:S3_secular_equation}
\end{equation}
Every physical real root $k_n(\mu)\in(0,\pi)$ determines a real reference energy through Eq.~\eqref{eq:S3_energy_dispersion}.

For $\mu=0$, Eq.~\eqref{eq:S3_secular_equation} gives
\begin{equation}
k_n(0)=\frac{n\pi}{L+1},
\qquad n=1,\ldots,L.
\label{eq:S3_open_momenta}
\end{equation}
Combining Eq.~\eqref{eq:S3_transformed_mode} with the gauge transformation above reproduces the usual open-chain skin modes. 

For nonzero impurity-PBC strength $\mu$, however, every mode acquires its own momentum $k_n(\mu)$, so different candidate pairs must be evaluated using their own finite-$\mu$ energies and spatial wavefunctions. Repeating the same calculation for the other chain with the opposite sign of $\kappa$, the right eigenmodes of both chains can be written uniformly as
\begin{equation}
\begin{aligned}
\langle x|\psi_{n,\ell}^{R}(\mu)\rangle
={}&
\mathcal N_{n,\ell}^{R}
e^{\eta_\ell\kappa(x-1)}
\Big[
\sin\!\big((L+1-x)k_n(\mu)\big)
\\
&\hspace{20mm}+
\mu e^{-\eta_\ell\kappa L}
\sin\!\big((x-1)k_n(\mu)\big)
\Big],
\end{aligned}
\label{eq:S3_exact_right_modes}
\end{equation}
where $\eta_a=+1$ and $\eta_b=-1$. Both chains have the same real energies $E_n(\mu)=2\cos k_n(\mu)$, with $k_n(\mu)$ determined by Eq.~\eqref{eq:S3_secular_equation}.

We order the real reference energies as
\begin{equation}
E_1(\mu)>E_2(\mu)>\cdots>E_L(\mu).
\label{eq:S3_energy_order}
\end{equation}
For each chain, define the right-eigenvector matrix
\begin{equation}
\Psi_\ell^R(\mu)=\left(|\psi_{1,\ell}^R\rangle,\ldots,|\psi_{L,\ell}^R\rangle\right),
\qquad \ell=a,b.
\end{equation}
The corresponding left eigenvectors are the rows of
\begin{equation}
\Psi_\ell^L(\mu)=\left[\Psi_\ell^R(\mu)\right]^{-1},
\label{eq:S3_left_inverse}
\end{equation}
and satisfy
\begin{equation}
\langle\psi_{i,\ell}^L|\psi_{j,\ell}^R\rangle=\delta_{ij}.
\label{eq:S3_biorthogonality}
\end{equation}

\subsection{Single impurity-chain real-to-complex threshold $\mu_{c}^{(0)}$}

The single-chain spectrum remains real while all physical roots of Eq.~\eqref{eq:S3_secular_equation} are real and nondegenerate. Its first real-to-complex transition occurs when two real roots merge into a double root. Requiring the secular equation and its $k$ derivative to vanish simultaneously is the standard condition for an isolated one-dimensional level coalescence (EP); it applies for arbitrary finite $L$ and $\kappa$ and uses no strong-$\mu$ approximation. The general finite-size threshold is therefore
\begin{equation}
\begin{aligned}
\mu_{c}^{(0)}
=
\inf\Big\{
\mu>0:\;&
F_L(k_c,\mu)=0,\qquad
\partial_kF_L(k_c,\mu)=0
\ \text{for some}\
k_c\in(0,\pi)
\Big\}.
\end{aligned}
\label{eq:S3_muc0_definition}
\end{equation}
The derivative entering this condition is
\begin{equation}
\begin{aligned}
\partial_kF_L(k,\mu)
={}&
(L+1)\cos[(L+1)k]
-
\mu^2(L-1)\cos[(L-1)k]
-
2\mu\cosh(\kappa L)\cos k.
\end{aligned}
\label{eq:S3_secular_derivative}
\end{equation}
Equations~\eqref{eq:S3_muc0_definition} and \eqref{eq:S3_secular_derivative}
together determine $\mu_{c}^{(0)}$ for arbitrary finite $L$, \emph{without} the use of any strong-$\mu$ approximation.

\begin{figure}[!htp]
    \centering
    \includegraphics[width=0.9\linewidth]{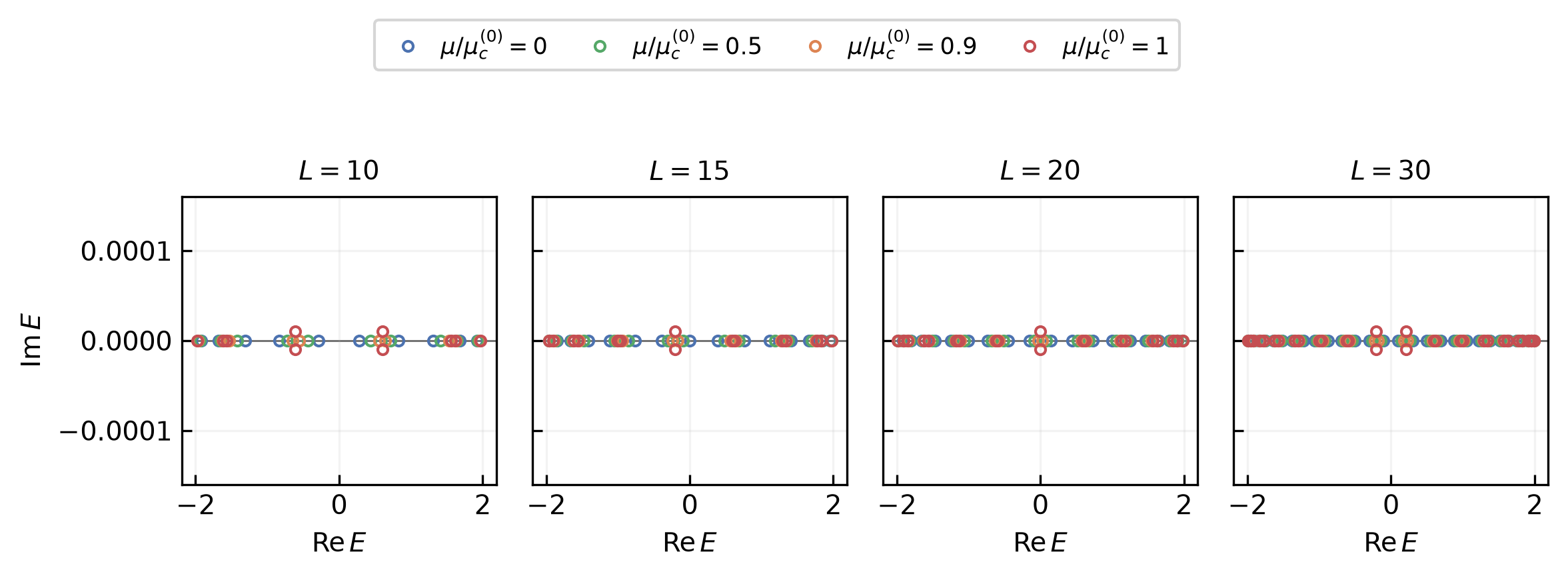}
    \caption{Isolated impurity-chain real-to-complex transition at different $L$ and impurity strength $\mu$. Panels (a)--(d) show the exact complex-energy spectrum of the single-chain Hamiltonian $H_b(\mu)$ in Eq.~\eqref{eq:S3_single_chain_model} for $L=10$, $15$, $20$, and $30$, respectively, with $\kappa=0.02$. Different colors correspond to $\mu/\mu_c^{(0)}=0$, $0.5$, $0.9$, and $1$, as indicated in the legend. Here $\mu_c^{(0)}$ is the first isolated-chain real-to-complex threshold defined by the double-root conditions in Eqs.~\eqref{eq:S3_muc0_definition} and \eqref{eq:S3_secular_derivative}. For $\mu<\mu_c^{(0)}$, all roots of the secular equation \eqref{eq:S3_secular_equation} remain real and the spectrum lies on the real axis. As $\mu$ approaches $\mu_c^{(0)}$, the first pair of real eigenenergies coalesces at an exceptional point. The location of this first coalescence depends on the system size; in particular, for $L=20=4\ell$ it occurs at $E=0$ and $\mu_c^{(0)}=e^{-\kappa L}$ according to Eq.~\eqref{eq:S3_muc0_L4ell}.}
    \label{fig:S3_single_chain_spectra}
\end{figure}

As the impurity-PBC strength $\mu$ increases, the corresponding evolution of the exact single-chain spectrum is shown in Fig.~\ref{fig:S3_single_chain_spectra}. For all $\mu<\mu_c^{(0)}$, the eigenenergies remain on the real axis, while the levels that generate the first single-chain exceptional point approach one another as $\mu\rightarrow\mu_c^{(0)}$. The location of this coalescing pair depends on $L$, demonstrating that the terminal reference-mode pair relevant near $\mu_c^{(0)}$ is system-size dependent.

The location of the first double root depends on the parity of $L/2$. For $L=4\ell$, one may set $k_c=\pi/2$, for which
\begin{equation}
F_L\left(\frac{\pi}{2},\mu\right)
=
1+\mu^2-2\mu\cosh(\kappa L).
\end{equation}
The first positive solution is then
\begin{equation}
\mu_{c}^{(0)}=e^{-\kappa L},
\qquad
L=4\ell.
\label{eq:S3_muc0_L4ell}
\end{equation}
By contrast, for $L=4\ell+2$,
\begin{equation}
F_L\left(\frac{\pi}{2},\mu\right)
=
-1-\mu^2-2\mu\cosh(\kappa L)<0
\end{equation}
for every positive $\mu$. 

For $\mu>\mu_{c}^{(0)}$, the reference chain is already complex at $\lambda=0$. Therefore the positive $\lambda$-driven real-to-complex threshold discussed below is defined only for $\mu<\mu_{c}^{(0)}$, with
\begin{equation}
\lambda_c(\mu)=0,
\qquad
\mu\geq\mu_{c}^{(0)}.
\label{eq:S3_lambda_above_muc0}
\end{equation}

\subsection{Two-level projection and pair-resolved impurity-closed threshold}

Consider a candidate pair $(i,j)$ of real single-chain reference levels. Since each level has one counterpart on each chain, the two-level projection spans the four-dimensional subspace
$\{|\psi_{i,b}^{R}\rangle,|\psi_{j,b}^{R}\rangle,
|\psi_{i,a}^{R}\rangle,|\psi_{j,a}^{R}\rangle\}$,
together with the corresponding biorthogonal left duals. This four-dimensional representation subsequently decomposes into two independent $2\times2$ blocks through the symmetric and antisymmetric combinations of the chain components.

Choosing the paired normalization of the two chains, direct projection of the physical inter-chain coupling gives
\begin{equation}
\begin{aligned}
v_{mn}(\mu)
&=
\langle\psi_{m,b}^{L}(\mu)|\mathcal V_{\perp}|\psi_{n,a}^{R}(\mu)\rangle
\\
&=
\langle\psi_{m,a}^{L}(\mu)|\mathcal V_{\perp}|\psi_{n,b}^{R}(\mu)\rangle,
\qquad m,n\in\{i,j\}.
\end{aligned}
\label{eq:S3_projected_interchain}
\end{equation}
The equality of the two directional projections follows from the explicit paired eigenmodes in Eq.~\eqref{eq:S3_exact_right_modes} and fixes a convenient common normalization convention.

In the ordered right basis
\begin{equation}
\mathcal B_{ij}^{R}=\left(|\psi_{i,b}^{R}\rangle,|\psi_{j,b}^{R}\rangle,|\psi_{i,a}^{R}\rangle,|\psi_{j,a}^{R}\rangle\right),
\end{equation}
the projected Hamiltonian is
\begin{equation}
\mathcal M_{ij}(\lambda,\mu)=
\begin{pmatrix}
E_i&0&\lambda v_{ii}&\lambda v_{ij}\\
0&E_j&\lambda v_{ji}&\lambda v_{jj}\\
\lambda v_{ii}&\lambda v_{ij}&E_i&0\\
\lambda v_{ji}&\lambda v_{jj}&0&E_j
\end{pmatrix}.
\label{eq:S3_four_mode_block}
\end{equation}
Equivalently, defining
\begin{equation}
\mathsf E_{ij}=
\begin{pmatrix}
E_i&0\\
0&E_j
\end{pmatrix},
\qquad
\mathsf V_{ij}=
\begin{pmatrix}
v_{ii}&v_{ij}\\
v_{ji}&v_{jj}
\end{pmatrix},
\end{equation}
Eq.~\eqref{eq:S3_four_mode_block} becomes
\begin{equation}
\mathcal M_{ij}(\lambda,\mu)=
\begin{pmatrix}
\mathsf E_{ij}&\lambda\mathsf V_{ij}\\
\lambda\mathsf V_{ij}&\mathsf E_{ij}
\end{pmatrix}.
\end{equation}
Equal and opposite linear combinations of the two chain components block-diagonalize this matrix as
\begin{equation}
\mathcal M_{ij}\sim\mathcal M_{ij}^{(+)}\oplus\mathcal M_{ij}^{(-)},
\end{equation}
with
\begin{equation}
\mathcal M_{ij}^{(s)}(\lambda,\mu)=
\begin{pmatrix}
E_i+s\lambda v_{ii}&s\lambda v_{ij}\\
s\lambda v_{ji}&E_j+s\lambda v_{jj}
\end{pmatrix},
\qquad s=\pm1.
\label{eq:S3_reduced_blocks}
\end{equation}
This reduction is performed entirely within the projected four-mode space and uses only matrix elements of the physical inter-chain coupling.

Define
\begin{equation}
\Delta E_{ij}(\mu)=E_i(\mu)-E_j(\mu),
\qquad
\Delta v_{ij}(\mu)=v_{ii}(\mu)-v_{jj}(\mu).
\label{eq:S3_active_definitions}
\end{equation}
The discriminant of the block labelled by $s$ is
\begin{equation}
\mathcal D_{ij}^{(s)}(\lambda,\mu)=\left[\Delta E_{ij}(\mu)+s\lambda\Delta v_{ij}(\mu)\right]^2+4\lambda^2v_{ij}(\mu)v_{ji}(\mu).
\label{eq:S3_pair_discriminant}
\end{equation}
A two-mode real-to-complex transition requires
\begin{equation}
v_{ij}(\mu)v_{ji}(\mu)<0.
\label{eq:S3_negative_coupling_product}
\end{equation}
Introducing
\begin{equation}
G_{ij}(\mu)=\sqrt{-v_{ij}(\mu)v_{ji}(\mu)},
\label{eq:S3_Gpq}
\end{equation}
the candidate exceptional-point roots are
\begin{equation}
\lambda_{c,\sigma}^{(i,j;s)}(\mu)=-\frac{\Delta E_{ij}(\mu)}{s\Delta v_{ij}(\mu)-2\sigma G_{ij}(\mu)},
\qquad s,\sigma=\pm1.
\label{eq:S3_candidate_roots}
\end{equation}
Because $\Delta E_{ij}>0$ for $i<j$, minimizing over $s$ and $\sigma$ gives the smallest positive root
\begin{equation}
\boxed{\lambda_c^{(i,j)}(\mu)=\frac{\Delta E_{ij}(\mu)}{|\Delta v_{ij}(\mu)|+2G_{ij}(\mu)}},
\qquad 0\leq\mu<\mu_c^{(0)}.
\label{eq:S3_pair_threshold}
\end{equation}
This derivation is exactly parallel to the OBC construction in Sec.~\ref{sec:S3_HN_selection}: $E_n$ and $v_{mn}$ are replaced by their impurity-dependent values $E_n(\mu)$ and $v_{mn}(\mu)$, while the four-mode-to-two-mode reduction, sign condition, and threshold formula retain the same structure. Thus $\mu$ enters only through the reference data; the EP condition itself is unchanged.
It is useful to define the projected activation strength
\begin{equation}
\Gamma_{ij}(\mu)=|\Delta v_{ij}(\mu)|+2G_{ij}(\mu),
\label{eq:S3_Gamma_pq}
\end{equation}
so that $\lambda_c^{(i,j)}(\mu)=\Delta E_{ij}(\mu)/\Gamma_{ij}(\mu)$. The finite-$\mu$ real-to-complex threshold of the coupled ladder, $\lambda_c(\mu)$, is the lower envelope
\begin{equation}
\lambda_c(\mu)=\min_{i<j}\lambda_c^{(i,j)}(\mu).
\label{eq:S3_global_threshold}
\end{equation}
The diagonal elements $v_{ii}$ and $v_{jj}$ and the product $v_{ij}v_{ji}$ are unchanged under reciprocal rescalings of each paired left-right eigenmode, so Eq.~\eqref{eq:S3_pair_threshold} is independent of the normalization convention.

\subsection{General sequence of impurity-driven activated-channel crossovers}

In general, as a parameter is varied, the mode pairs that form the lower envelope of Eq.~\eqref{eq:S3_global_threshold} can switch. Suppose the mode pairs that successively form the global lower envelope are
\begin{equation}
    (p_0,q_0),(p_1,q_1),\ldots,(p_M,q_M),
    \label{eq:S3_winning_sequence}
\end{equation}
and let the corresponding switching points be
\begin{equation}
    0<\mu_1<\mu_2<\cdots<\mu_M<\mu_{c}^{(0)}.
    \label{eq:S3_switch_sequence}
\end{equation}
The pair $(p_j,q_j)$ is the activated channel throughout the interval between the $j$th and $(j+1)$th switching points.

A physical crossover $\mu_j$ is not merely an intersection between two arbitrary pair-resolved threshold curves. It must lie on the global lower envelope and therefore satisfy
\begin{equation}
\begin{aligned}
    \lambda_c^{(p_{j-1},q_{j-1})}(\mu_j)
    &=
    \lambda_c^{(p_j,q_j)}(\mu_j)
    \\
    &=
    \min_{i<k}
    \lambda_c^{(i,k)}(\mu_j),
    \qquad
    j=1,\ldots,M.
\end{aligned}
\label{eq:S3_lower_envelope_intersection}
\end{equation}
Thus, intersections between pair-resolved curves that lie above another candidate threshold do not correspond to changes of the first activated channel.

Defining
\begin{equation}
    \mu_0=0,
    \qquad
    \mu_{M+1}=\mu_{c}^{(0)},
    \label{eq:S3_interval_endpoints}
\end{equation}
the physical threshold takes the explicit piecewise form
\begin{equation}
\lambda_c(\mu)
=
\begin{cases}
    \lambda_c^{(p_0,q_0)}(\mu),
    &
    0\leq\mu<\mu_1,
    \\[1mm]
    \lambda_c^{(p_1,q_1)}(\mu),
    &
    \mu_1\leq\mu<\mu_2,
    \\
    \vdots
    &
    \vdots
    \\[-1mm]
    \lambda_c^{(p_M,q_M)}(\mu),
    &
    \mu_M\leq\mu<\mu_{c}^{(0)},
    \\[1mm]
    0,
    &
    \mu\geq\mu_{c}^{(0)}.
\end{cases}
\label{eq:S3_general_piecewise}
\end{equation}
At each switching point $\mu_j$, the two neighboring expressions in Eq.~\eqref{eq:S3_general_piecewise} give the same value by Eq.~\eqref{eq:S3_lower_envelope_intersection}. Assigning the equality to either neighboring interval is therefore only a convention.

Equivalently, the result may be written compactly as
\begin{equation}
    \lambda_c(\mu)
    =
    \lambda_c^{(p_j,q_j)}(\mu),
    \qquad
    \mu_j\leq\mu<\mu_{j+1},
    \qquad
    j=0,\ldots,M.
    \label{eq:S3_general_piecewise_compact}
\end{equation}

The number $M$ of switching points is not fixed a priori. Depending on $L$ and $\kappa$, the lower envelope may contain no crossover, a single crossover, or several successive crossovers. In particular, the channel selected at $\mu=0$ need not switch directly to the pair that becomes defective at $\mu_{c}^{(0)}$; one or more intermediate mode pairs may form the lower envelope over finite intervals of $\mu$.

For a real and nondegenerate spectrum immediately below the first isolated second-order exceptional point, the two colliding eigenenergy branches must be adjacent in the ordered reference spectrum. Therefore, when the modes are ordered according to Eq.~\eqref{eq:S3_energy_order}, the winning pairs satisfy
\begin{equation}
    q_j=p_j+1,
    \qquad
    j=0,\ldots,M.
    \label{eq:S3_adjacent_winning_pairs}
\end{equation}
For the present coupled HN ladder specifically, the combined operation of spatial reflection $x\mapsto L+1-x$, chain exchange $a\leftrightarrow b$, and reversal $\kappa\mapsto-\kappa$ enforces spectral reflection. Consequently, the reflected pair
\begin{equation}
    (L+1-q_j,L+1-p_j)
\end{equation}
has the same threshold. This degeneracy is a model-specific symmetry property, not a general consequence of activated-mode theory. We use one canonical representative of each reflected pair in the examples below.

\subsection{Two impurity-closed examples}

\subsubsection{Example I: three-channel sequence $(1,2)\to(2,3)\to(4,5)$ for $L=10$}

We first take
\begin{equation}
L=10,
\qquad
\kappa=0.02.
\label{eq:S3_L10_parameters}
\end{equation}
Since $L=4\ell+2$, $\ell$ an integer, the first single-chain collision does not occur at $k=\pi/2$. Solving Eqs.~\eqref{eq:S3_muc0_definition} and \eqref{eq:S3_secular_derivative} gives
\begin{equation}
\begin{aligned}
\mu_{c}^{(0)}
&=
0.83,
\\
k_c
&=
1.262
=
0.40\pi,
\\
E_c
&=
2\cos k_c
=
0.61.
\end{aligned}
\label{eq:S3_L10_muc0}
\end{equation}
By spectral reflection, an equivalent collision occurs simultaneously at energy $-E_c$. In the canonical mode labeling, the reference pairs approaching these two exceptional points are
\begin{equation}
(4,5)
\qquad\text{and}\qquad
(6,7).
\end{equation}

The lower envelope exhibits two distinct channel crossovers. The first is
\begin{equation}
\begin{aligned}
\mu_1
&=
0.25,
\\
\frac{\mu_1}{\mu_{c}^{(0)}}
&=
0.31,
\end{aligned}
\label{eq:S3_L10_mu1}
\end{equation}
where
\begin{equation}
\begin{aligned}
\lambda_c(\mu_1)
&=
\lambda_c^{(2,3)}(\mu_1)
=
0.13.
\end{aligned}
\label{eq:S3_L10_lambda1}
\end{equation}
The winning pair changes from $(1,2)$ to $(2,3)$.

The second crossover occurs at
\begin{equation}
\begin{aligned}
\mu_2
&=
0.57,
\\
\frac{\mu_2}{\mu_{c}^{(0)}}
&=
0.69,
\end{aligned}
\label{eq:S3_L10_mu2}
\end{equation}
where
\begin{equation}
\begin{aligned}
\lambda_c^{(2,3)}(\mu_2)
&=
\lambda_c^{(4,5)}(\mu_2)
=
0.066.
\end{aligned}
\label{eq:S3_L10_lambda2}
\end{equation}
The active channel then changes from $(2,3)$ to the pair $(4,5)$ that approaches the positive-energy single-chain exceptional point.

\begin{figure}[!htp]
    \centering
    \subfigure[]{\includegraphics[width=0.23\linewidth]{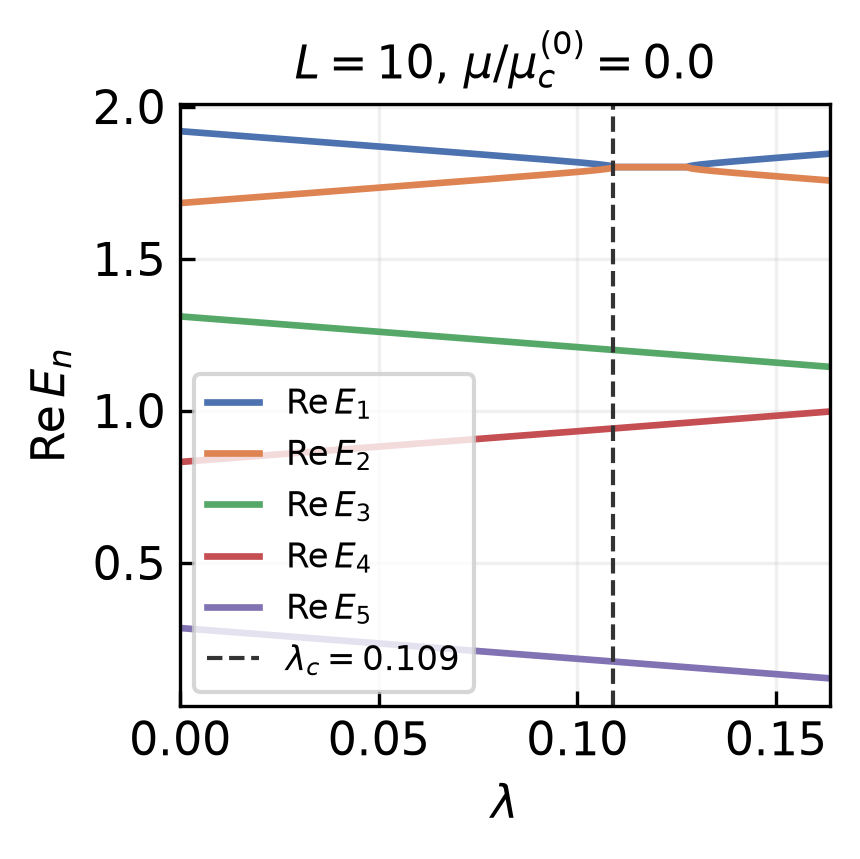}}
    \subfigure[]{\includegraphics[width=0.23\linewidth]{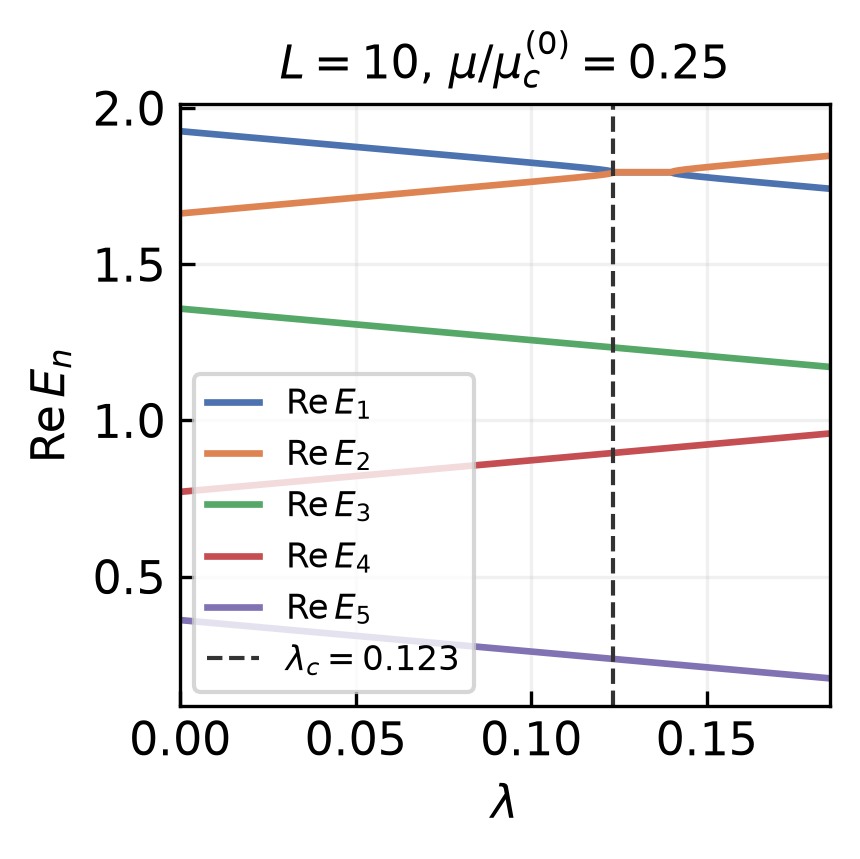}
    }
    \subfigure[]{\includegraphics[width=0.23\linewidth]{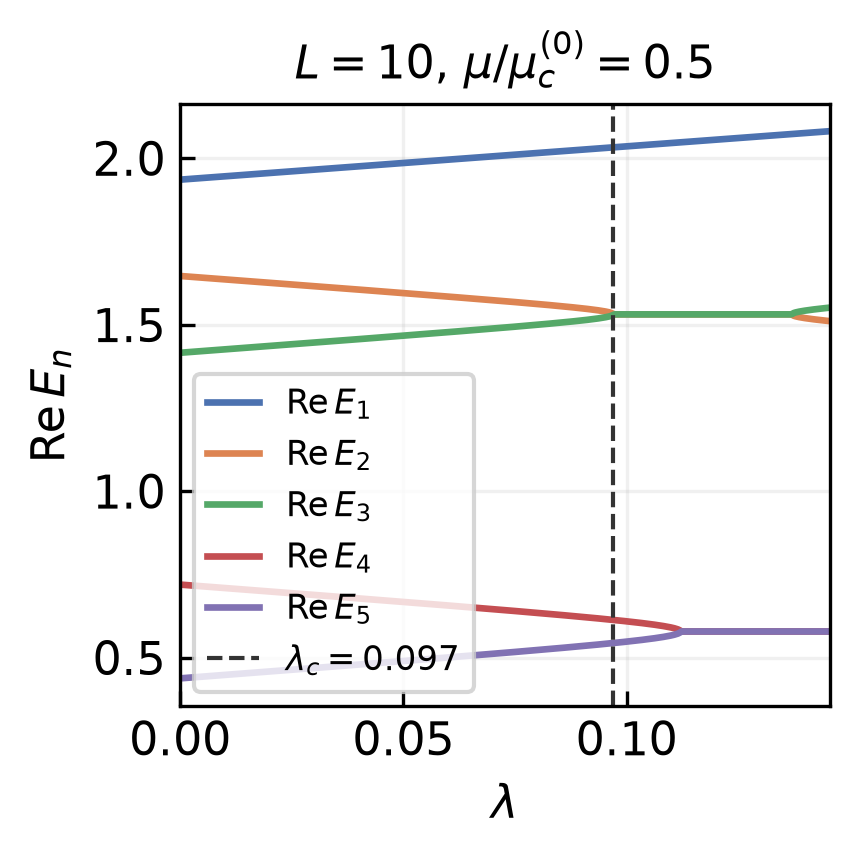}
    }
    \subfigure[]{\includegraphics[width=0.23\linewidth]{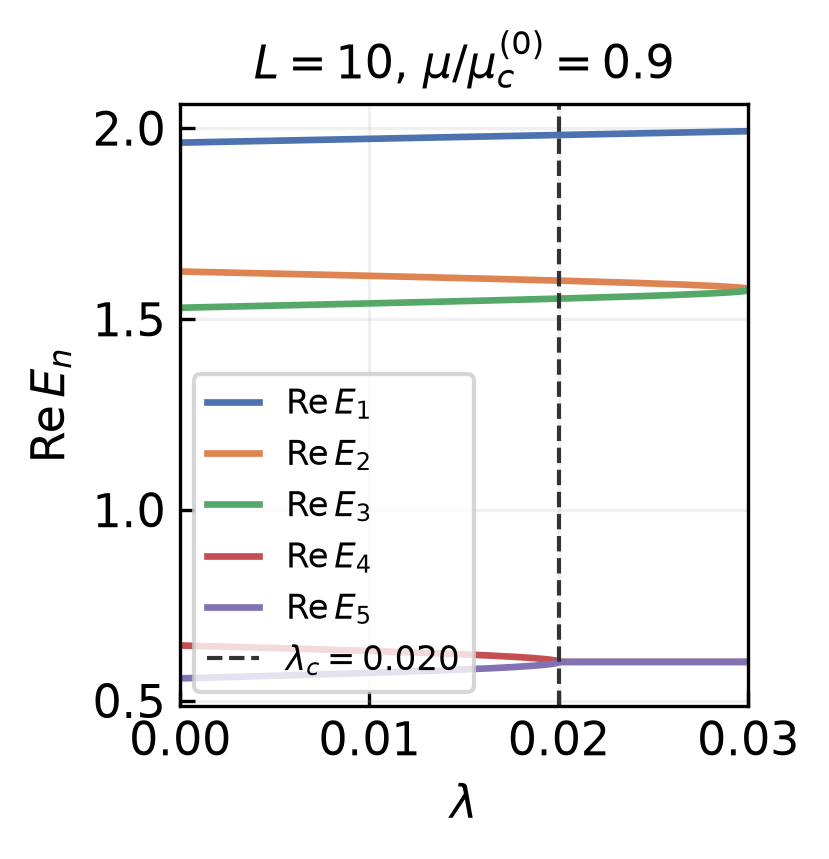}
    }
    \subfigure[]{\includegraphics[width=0.23\linewidth]{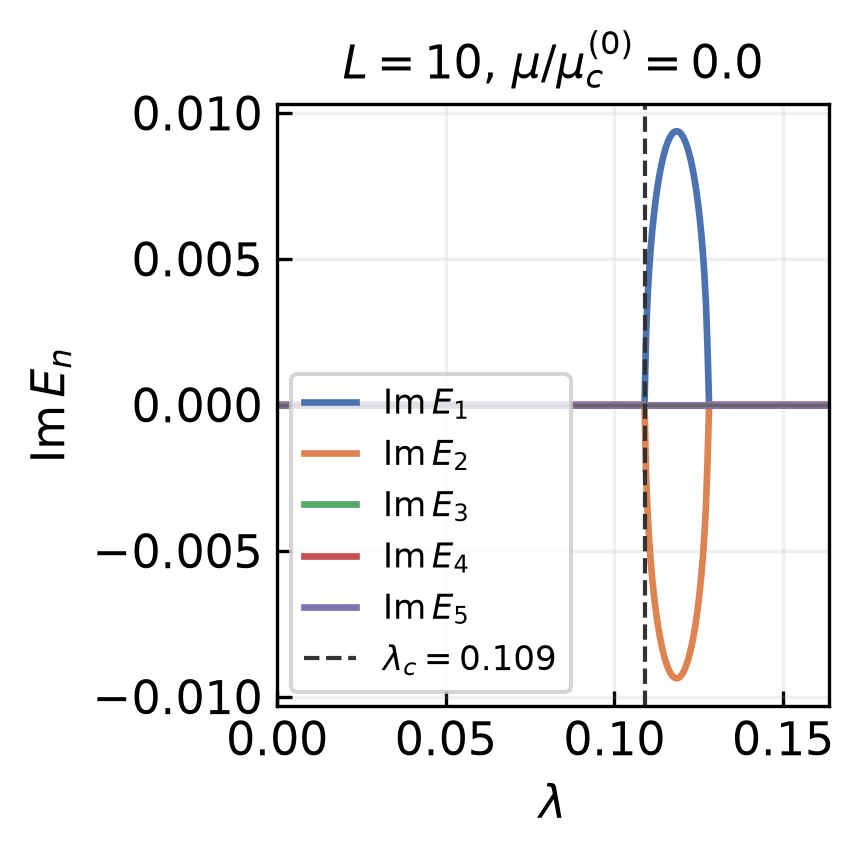}
    }
    \subfigure[]{\includegraphics[width=0.23\linewidth]{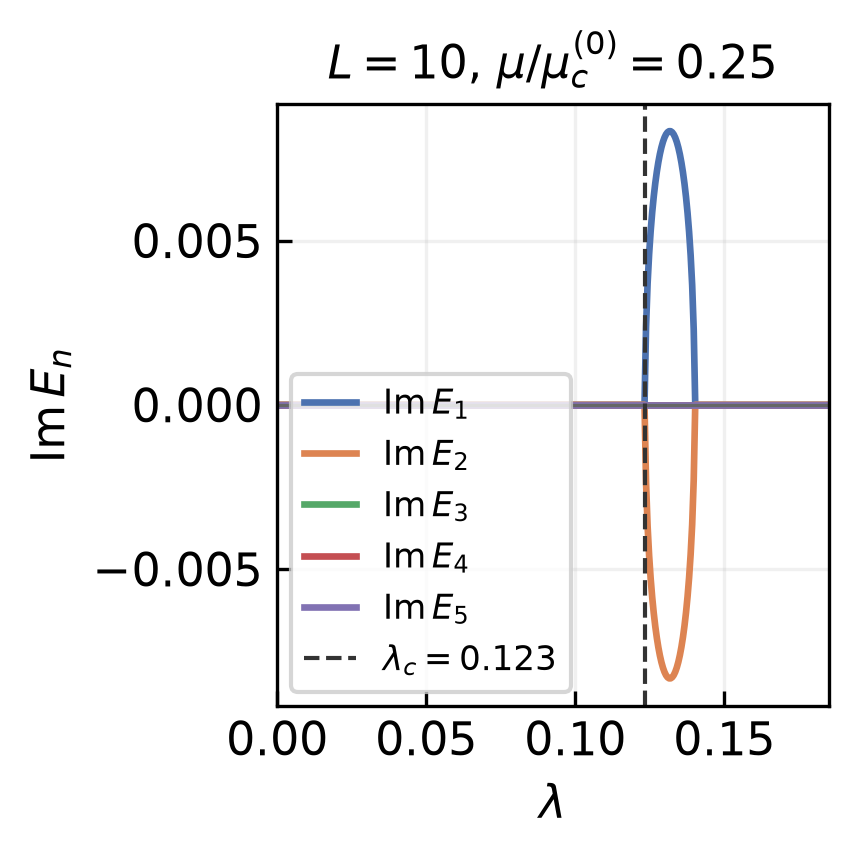}
    }
    \subfigure[]{\includegraphics[width=0.23\linewidth]{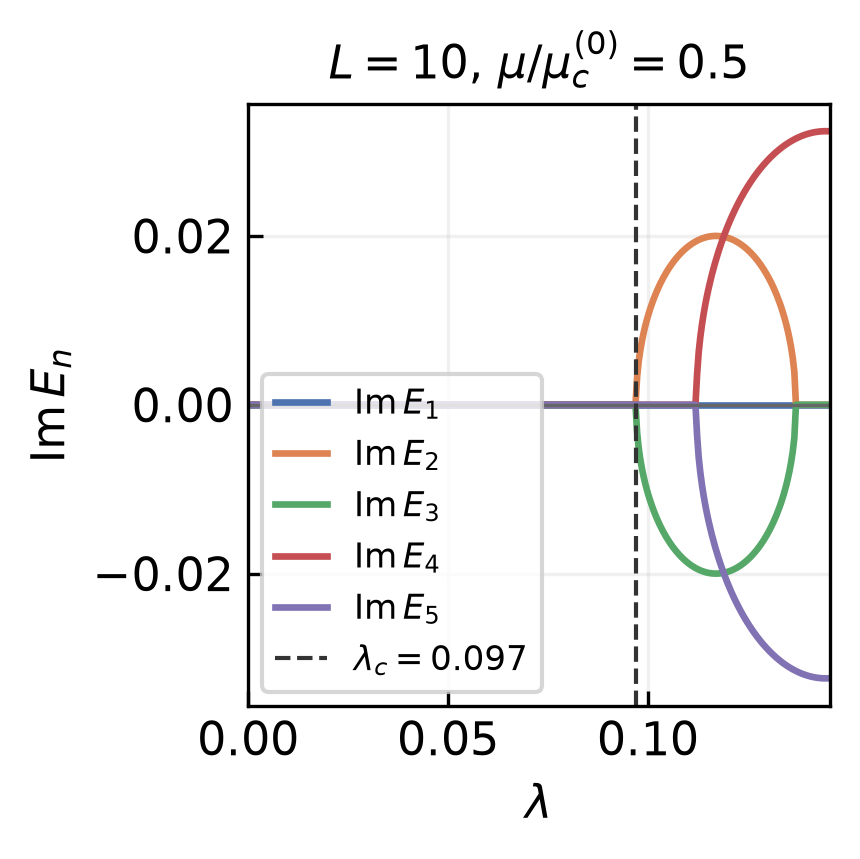}
    }
    \subfigure[]{\includegraphics[width=0.23\linewidth]{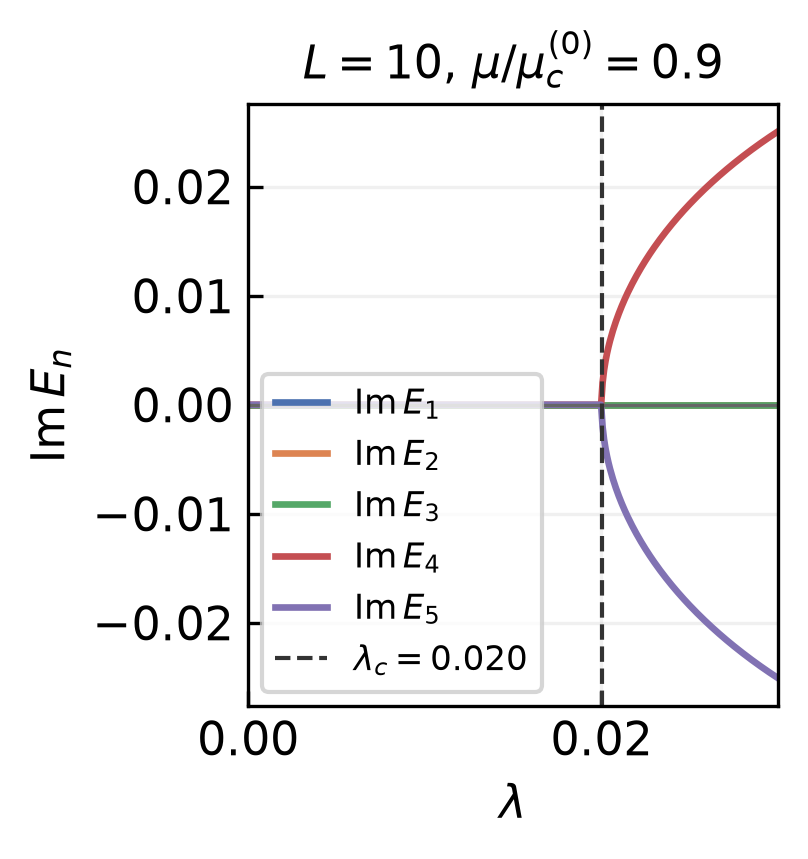}
    }
    \caption{Branch-resolved spectral evolution of the full coupled Hatano-Nelson ladder in Eq.~\eqref{eq:S3_full_model} for $L=10$ and $\kappa=0.02$. Columns correspond to $\mu/\mu_c^{(0)}=0$, $0.25$, $0.5$, and $0.9$. The upper and lower rows show, respectively, $\operatorname{Re}E_n(\lambda)$ and $\operatorname{Im}E_n(\lambda)$ for the continuously tracked reference branches $n=1,\ldots,5$, matched to the ordered single-chain levels at $\lambda=0$. In each panel, the vertical dashed line marks the first real-to-complex threshold $\lambda_c(\mu)$ extracted from the complete $2L\times2L$ ladder spectrum. This threshold is the lower envelope of the pair-resolved expression in Eq.~\eqref{eq:S3_pair_threshold}, as defined by Eq.~\eqref{eq:S3_global_threshold}. Below $\lambda_c$, the displayed branches remain real; at $\lambda_c$, the activated pair coalesces; and above it, the same pair develops complex-conjugate imaginary parts. The four columns sample the three regimes in Eq.~\eqref{eq:S3_L10_piecewise}: the activated channel is $(1,2)$ for $\mu/\mu_c^{(0)}=0$ and $0.25$, $(2,3)$ for $\mu/\mu_c^{(0)}=0.5$, and $(4,5)$ for $\mu/\mu_c^{(0)}=0.9$, directly realizing the sequence in Eq.~\eqref{eq:S3_L10_sequence}.}
    \label{fig:S3_ReIm_lambda_L10}
\end{figure}

Writing
\begin{equation}
r=\frac{\mu}{\mu_{c}^{(0)}},
\end{equation}
the complete piecewise threshold is
\begin{equation}
\lambda_c(\mu)
=
\begin{cases}
\lambda_c^{(1,2)}(\mu),
&
0\leq r<0.31, \\[1mm]
\lambda_c^{(2,3)}(\mu),
&
0.31
<r<
0.69, \\[1mm]
\lambda_c^{(4,5)}(\mu),
&
0.69
<r<1, \\[1mm]
0,
&
r\geq1.
\end{cases}
\label{eq:S3_L10_piecewise}
\end{equation}
Thus the activated-channel sequence is
\begin{equation}
(1,2)
\longrightarrow
(2,3)
\longrightarrow
(4,5).
\label{eq:S3_L10_sequence}
\end{equation}

The corresponding branch-resolved spectral flow is shown in Fig.~\ref{fig:S3_ReIm_lambda_L10}. At small $\mu$, the first collision occurs between the continuously tracked band-edge branches $(1,2)$. After the first lower-envelope crossover, the earliest collision moves to $(2,3)$, and after the second crossover it moves to $(4,5)$. The continuous evolution of all displayed branches demonstrates that these crossovers are genuine changes of the first exceptional-point channel, rather than discontinuous relabelings of the reference eigenenergies.

The reflected partners $(9,10)$, $(8,9)$, and $(6,7)$ have identical pair-resolved thresholds. Equation~\eqref{eq:S3_L10_sequence} uses the canonical representative of each reflected pair.

This example demonstrates that the crossover need not be a direct handoff between the open-chain band-edge channel and the final single-chain-EP channel. The intermediate pair $(2,3)$ forms the lower envelope over a finite interval of $\mu$.

\subsubsection{Example II: direct sequence $(1,2)\to(14,15)$ for $L=30$}

\begin{figure}
    \centering
    \subfigure[]{\includegraphics[width=0.23\linewidth]{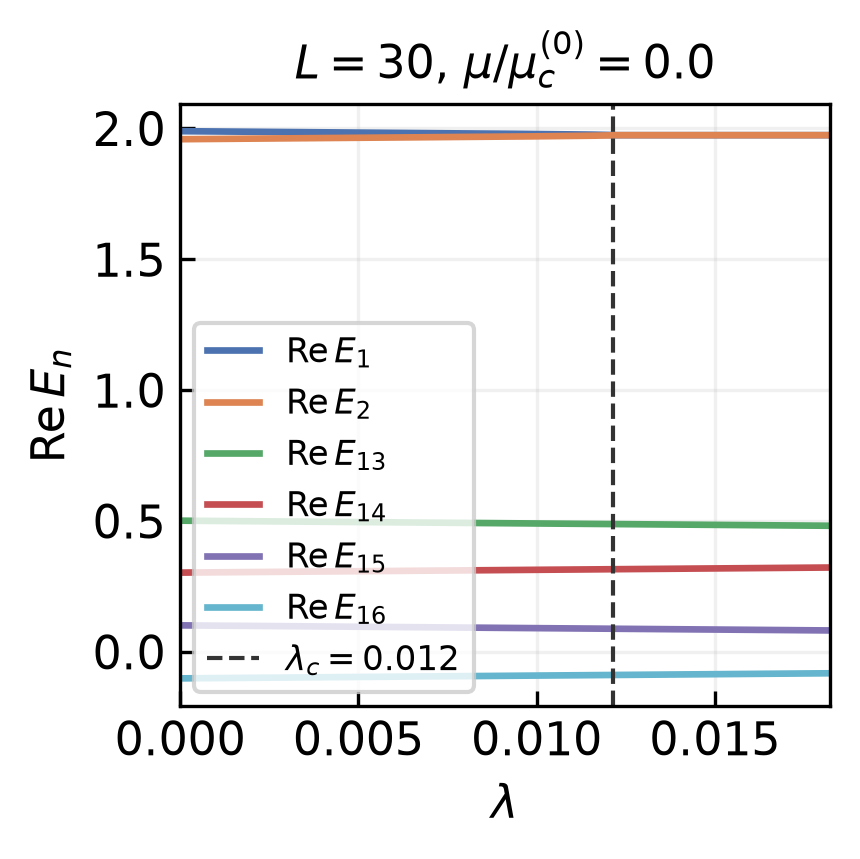}}
    \subfigure[]{\includegraphics[width=0.23\linewidth]{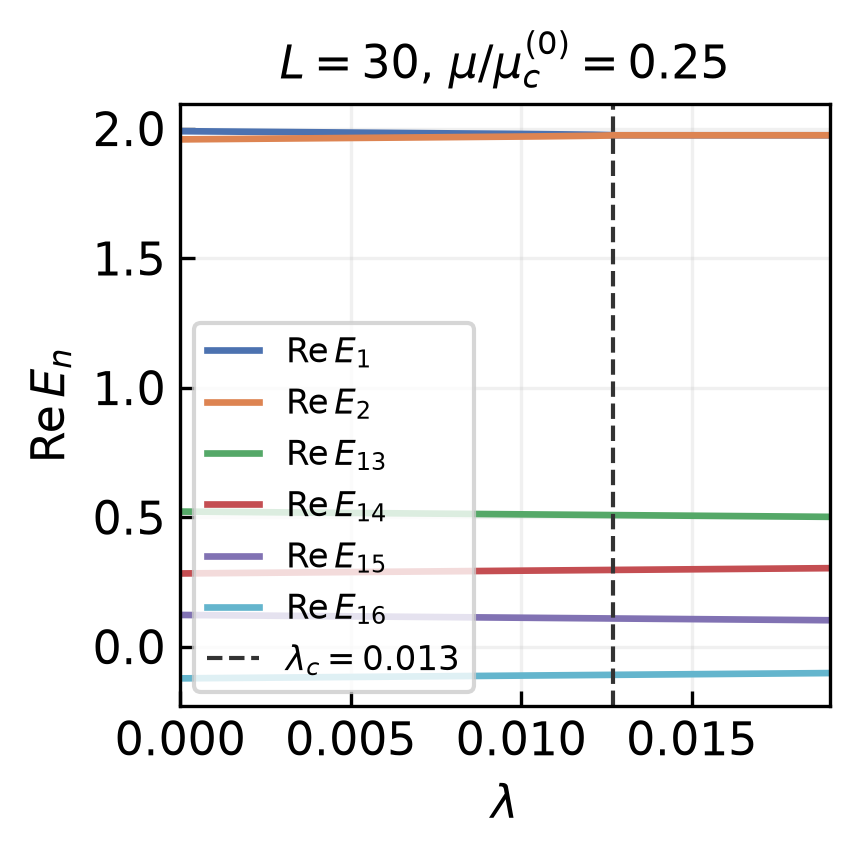}}
    \subfigure[]{\includegraphics[width=0.23\linewidth]{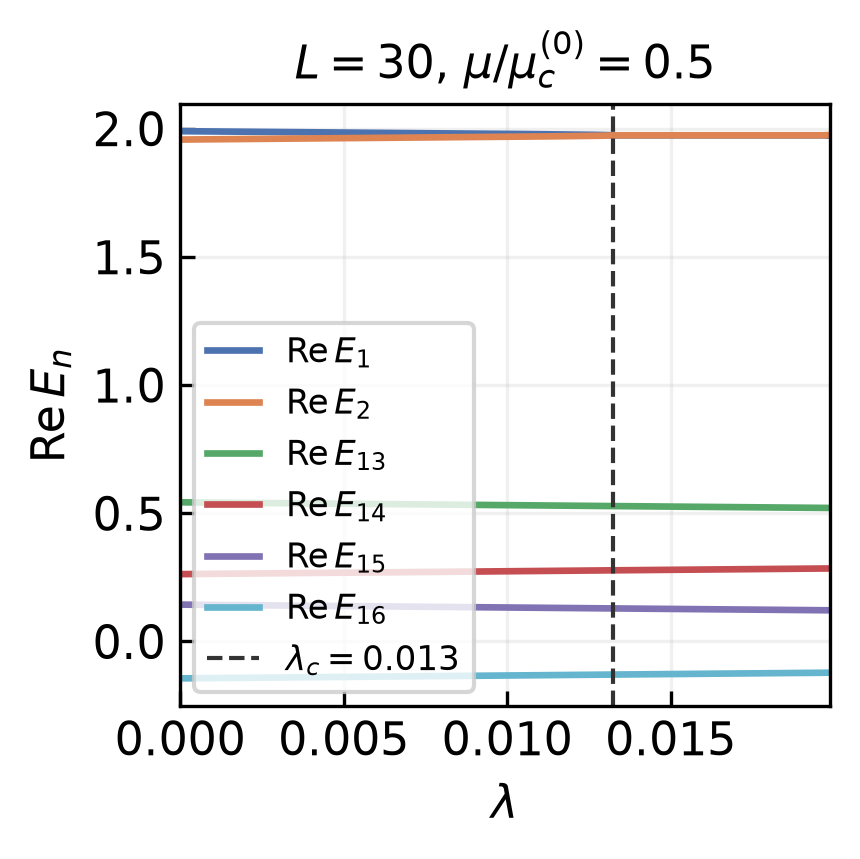}}
    \subfigure[]{\includegraphics[width=0.23\linewidth]{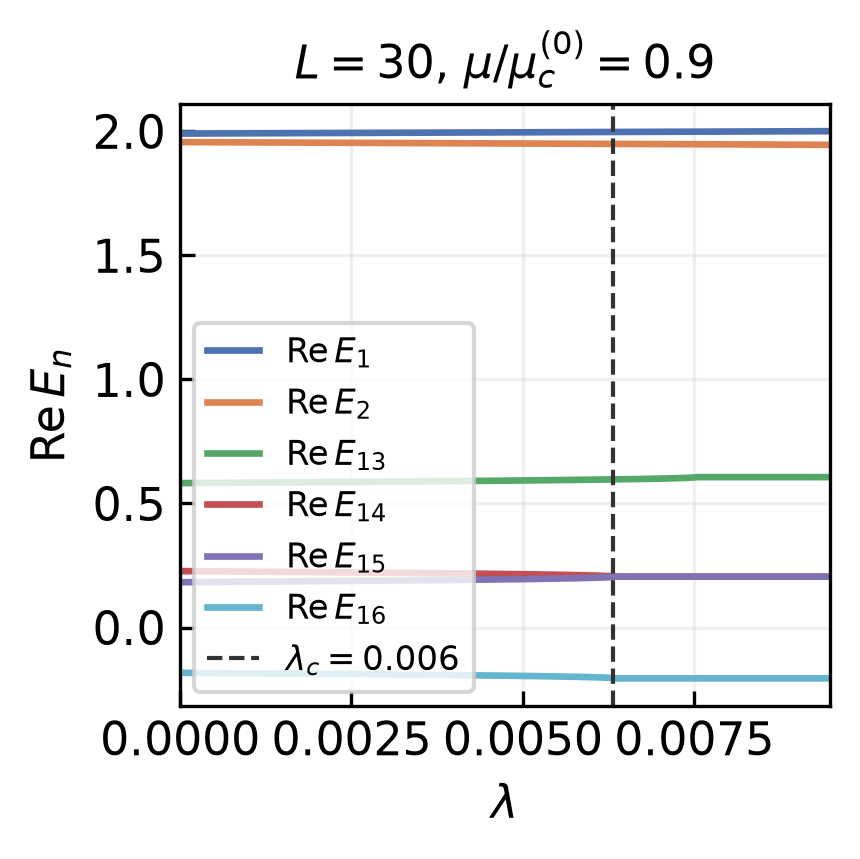}}
    \subfigure[]{\includegraphics[width=0.23\linewidth]{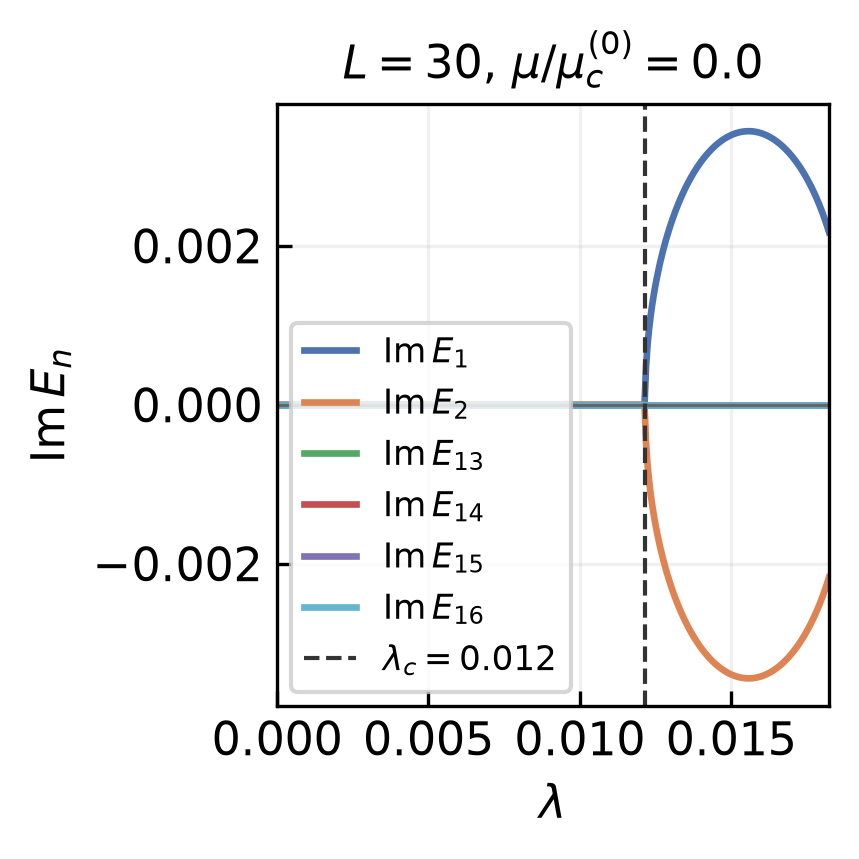}}
    \subfigure[]{\includegraphics[width=0.23\linewidth]{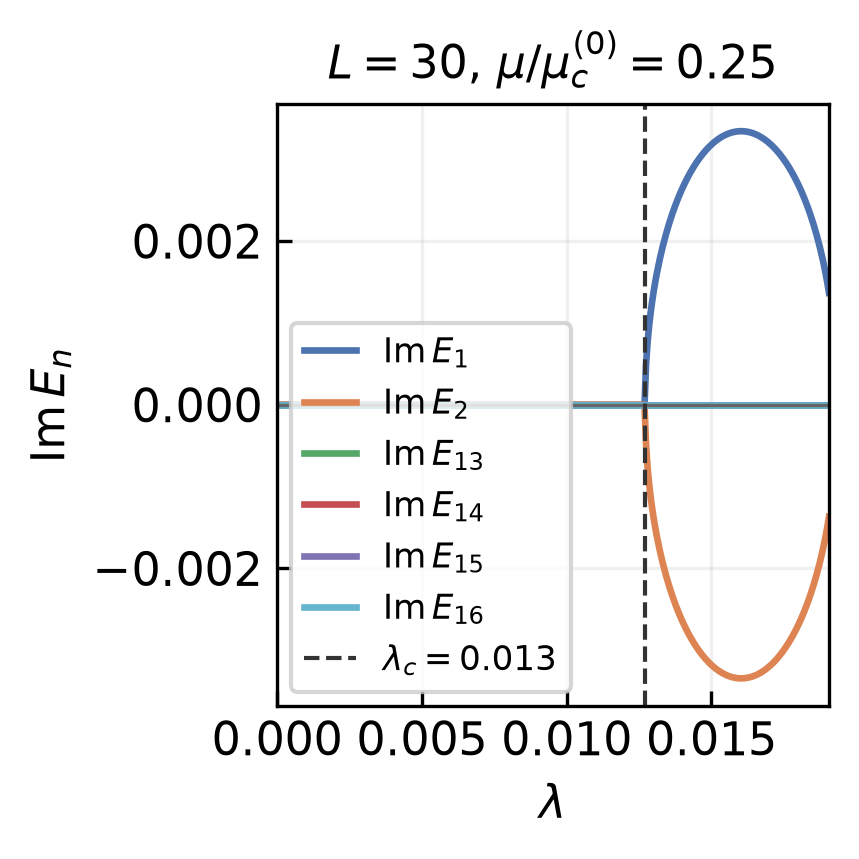}}
    \subfigure[]{\includegraphics[width=0.23\linewidth]{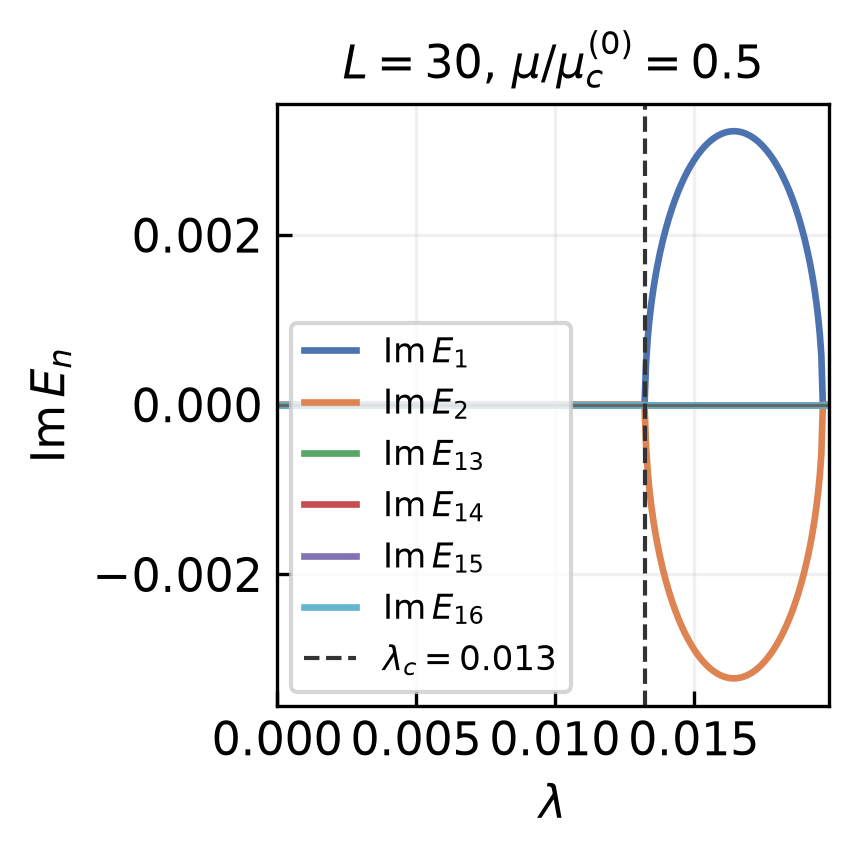}}
    \subfigure[]{\includegraphics[width=0.23\linewidth]{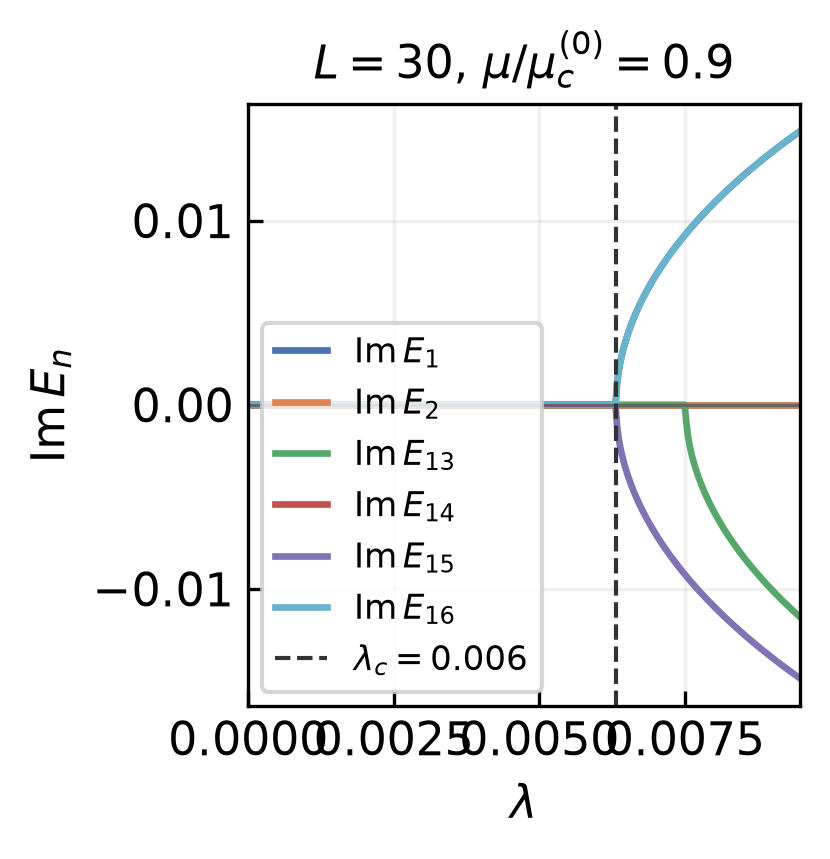}}
    \caption{Branch-resolved spectral evolution of the full coupled Hatano-Nelson ladder in Eq.~\eqref{eq:S3_full_model} for $L=30$ and $\kappa=0.02$. Columns correspond to $\mu/\mu_c^{(0)}=0$, $0.25$, $0.5$, and $0.9$. The upper row shows $\operatorname{Re}E_n(\lambda)$ and the lower row shows $\operatorname{Im}E_n(\lambda)$ for the continuously tracked reference branches $n=1,2,13,14,15,16$, matched to the ordered single-chain levels at $\lambda=0$. The threshold is extracted from the complete $2L\times2L$ ladder spectrum and is given by the global lower envelope in Eq.~\eqref{eq:S3_global_threshold} of the pair-resolved thresholds in Eq.~\eqref{eq:S3_pair_threshold}. The vertical dashed line in each panel marks the first real-to-complex transition, where the real parts of the activated branches coalesce and their imaginary parts bifurcate into a complex-conjugate pair. The first three columns satisfy $\mu/\mu_c^{(0)}<0.79$ and remain controlled by the band-edge pair $(1,2)$, whereas the column at $\mu/\mu_c^{(0)}=0.9$ lies beyond the switch and is controlled by the terminal pair $(14,15)$, as summarized by Eq.~\eqref{eq:S3_L30_sequence}.} 
    \label{fig:S3_ReIm_lambda_L30}
\end{figure}

Figure~\ref{fig:S3_ReIm_lambda_L30} shows that the continuously tracked band-edge branches $(1,2)$ remain the first pair to coalesce for $\mu/\mu_c^{(0)}=0$, $0.25$, and $0.5$. At $\mu/\mu_c^{(0)}=0.9$, which lies above the switching point $\mu_1/\mu_c^{(0)}\simeq0.79$, the first coalescence has moved to the terminal pair $(14,15)$. The figure therefore distinguishes continuous spectral deformation from the actual change of the global lower-envelope minimizer. 

We next take
\begin{equation}
L=30,
\qquad
\kappa=0.02.
\label{eq:S3_L30_parameters}
\end{equation}
Solving the single-chain double-root equations gives
\begin{equation}
\begin{aligned}
\mu_{c}^{(0)}
&=
0.55,
\\
k_c
&=
1.47
=
0.47\pi,
\\
E_c
&=
2\cos k_c
=
0.21.
\end{aligned}
\label{eq:S3_L30_muc0}
\end{equation}
The reflected exceptional point occurs at energy $-E_c$. The corresponding canonical terminal pair is $(14,15)$, with reflected partner $(16,17)$.

In this case, only one lower-envelope crossover occurs:
\begin{equation}
\begin{aligned}
\mu_1
&=
0.43,
\\
\frac{\mu_1}{\mu_{c}^{(0)}}
&=
0.79.
\end{aligned}
\label{eq:S3_L30_mu1}
\end{equation}
At the crossing,
\begin{equation}
\begin{aligned}
\lambda_c(\mu_1)
&=
\lambda_c^{(14,15)}(\mu_1)
\\
&=
0.014.
\end{aligned}
\label{eq:S3_L30_lambda1}
\end{equation}
The threshold is therefore
\begin{equation}
\lambda_c(\mu)
=
\begin{cases}
\lambda_c^{(1,2)}(\mu),
&
0\leq
\mu/\mu_{c}^{(0)}
<
0.79, \\[1mm]
\lambda_c^{(14,15)}(\mu),
&
0.79
<
\mu/\mu_{c}^{(0)}
<1, \\[1mm]
0,
&
\mu/\mu_{c}^{(0)}
\geq1.
\end{cases}
\label{eq:S3_L30_piecewise}
\end{equation}
The activated-channel sequence is simply
\begin{equation}
(1,2)
\longrightarrow
(14,15).
\label{eq:S3_L30_sequence}
\end{equation}
\begin{figure}[!htp]
\centering
\subfigure[$L=10$]{
\includegraphics[
width=0.47\linewidth
]{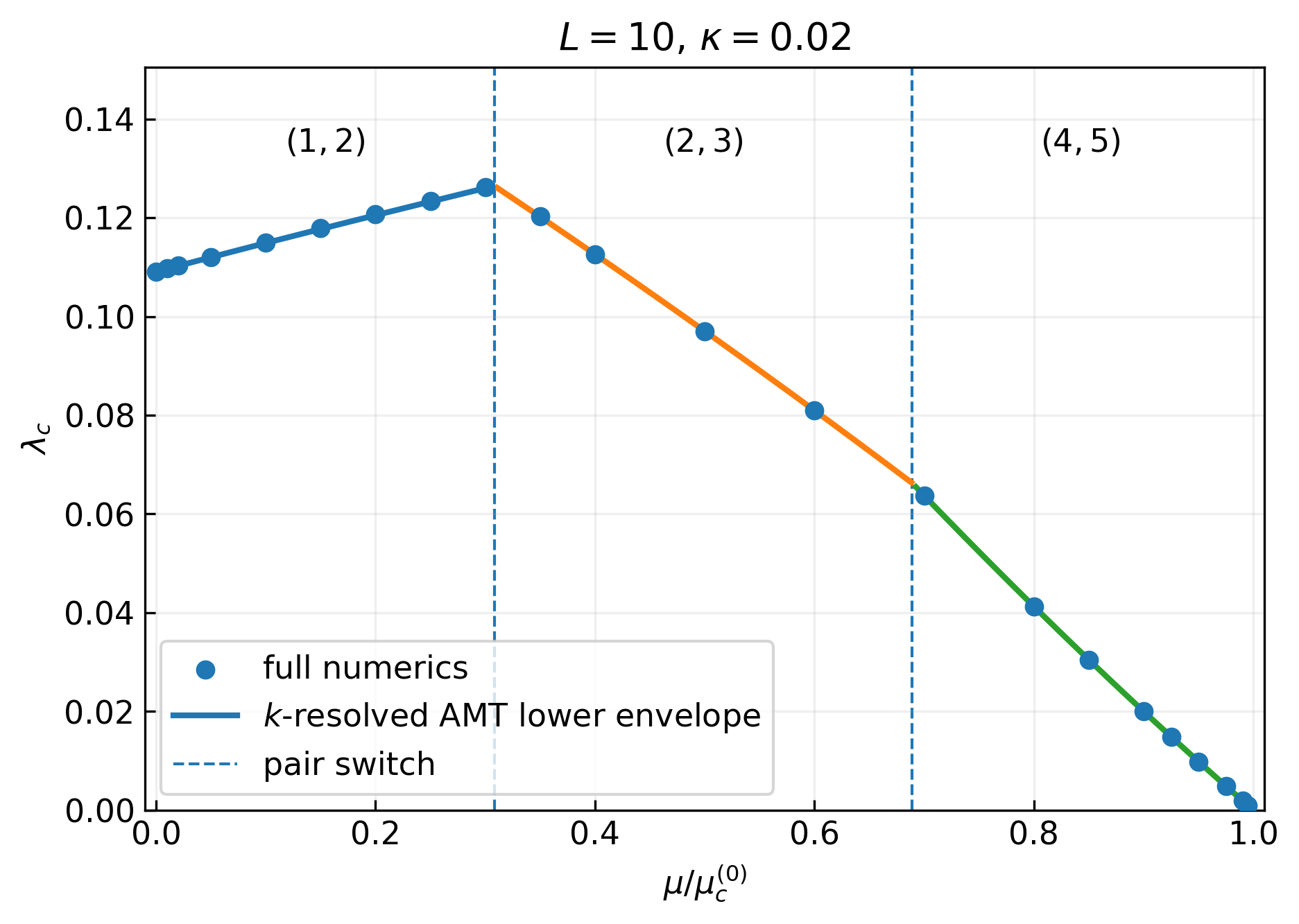}
}
\hfill
\subfigure[$L=30$]{
\includegraphics[
width=0.47\linewidth
]{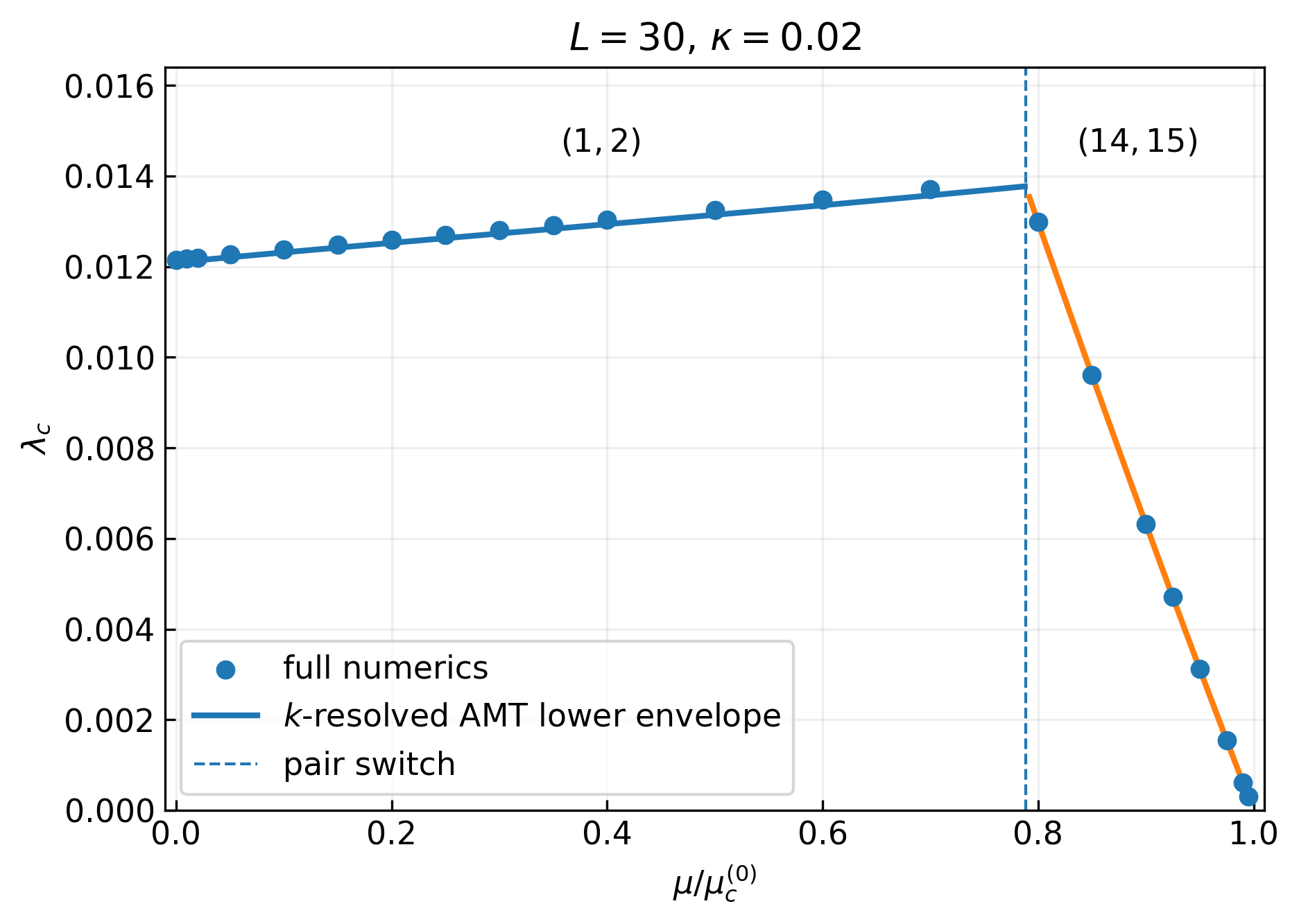}
}
\caption{
Finite-$\mu$ activated-channel crossovers for $\kappa=0.02$ in the unit-hopping convention of Eq.~\eqref{eq:S3_full_model}. Dots are thresholds extracted from the complete $2L\times2L$ spectrum, and solid curves are the pair-resolved activated-mode lower envelope. Dashed vertical lines mark the lower-envelope intersections in Eq.~\eqref{eq:S3_lower_envelope_intersection}. The pair labels indicate the canonical representative under spectral reflection. (a) For $L=10$, the threshold passes through the sequence $(1,2)\rightarrow(2,3)\rightarrow(4,5)$ and therefore contains two crossovers. (b) For $L=30$, the threshold switches directly from $(1,2)$ to $(14,15)$.
}
\label{fig:S3_multiple_crossovers}
\end{figure}

\subsection{Origin of the nonmonotonic impurity-closed threshold for $L=10$}
\label{sec:S3_L10_nonmonotonicity}

In the main text, Fig.~2(b2) showed that, for $L=10$, the threshold $\lambda_c(\mu)$ depends nonmonotonically on the boundary coupling: it first increases with $\mu$, and then decreases toward zero as $\mu\rightarrow\mu_c^{(0)}$, accompanied by successive changes of the activated channel. Below, we explain this behavior by separating the two competing mechanisms entering the pair-resolved threshold. At small $\mu$, the activated channel remains $(1,2)$, and the boundary coupling drives the two reference energies $E_1$ and $E_2$ farther apart, thereby increasing their spectral detuning and hence $\lambda_c$. At larger $\mu$, the global lower envelope switches to progressively more central mode pairs, eventually selecting the terminal pair whose detuning collapses at the isolated-chain exceptional point; this channel switching reverses the trend and forces $\lambda_c$ to vanish as $\mu\rightarrow\mu_c^{(0)}$.

The initial increase of the $L=10$ threshold is already determined within the first winning channel $(1,2)$. For $0\leq\mu<\mu_1$, Eq.~\eqref{eq:S3_pair_threshold} gives
\begin{equation}
    \lambda_c(\mu)
    =
    \lambda_c^{(1,2)}(\mu)
    =
    \frac{\Delta E_{12}(\mu)}{\Gamma_{12}(\mu)},
    \qquad
    \Gamma_{12}(\mu)
    =
    |\Delta v_{12}(\mu)|+2G_{12}(\mu).
    \label{eq:S3_L10_ratio}
\end{equation}
The numerator is the separation of the two upper band-edge reference levels. Its small-$\mu$ behavior follows directly from the exact secular equation~\eqref{eq:S3_secular_equation}. Let
\begin{equation}
    \theta=\frac{\pi}{L+1},
    \qquad
    C_L=\cosh(\kappa L),
\end{equation}
and write $k_n(\mu)=n\theta+\mu k_n'(0)+O(\mu^2)$. Implicit differentiation of
$F_L[k_n(\mu),\mu]=0$ at $\mu=0$ gives
\begin{equation}
    k_n'(0)
    =
    \frac{2C_L(-1)^n\sin(n\theta)}{L+1}.
    \label{eq:S3_small_mu_k_shift}
\end{equation}
Since $E_n(\mu)=2\cos k_n(\mu)$,
\begin{equation}
    E_n'(0)
    =
    -\frac{4C_L(-1)^n\sin^2(n\theta)}{L+1}.
    \label{eq:S3_small_mu_E_shift}
\end{equation}
In particular,
\begin{equation}
\begin{aligned}
    E_1(\mu)
    & =
    E_1(0)
    +
    \frac{4C_L\sin^2\theta}{L+1}\mu
    +O(\mu^2),
    \\
    E_2(\mu)
    & =
    E_2(0)
    -
    \frac{4C_L\sin^2(2\theta)}{L+1}\mu
    +O(\mu^2).
\end{aligned}
\label{eq:S3_small_mu_E12}
\end{equation}
Thus the boundary closure moves the two levels away from one another: $E_1$ increases while $E_2$ decreases. Their detuning therefore opens linearly,
\begin{equation}
\begin{aligned}
    \Delta E_{12}(\mu)
    ={}&
    2\left(\cos\theta-\cos2\theta\right)
    \\
    &+
    \frac{4C_L}{L+1}
    \left[\sin^2\theta+\sin^2(2\theta)\right]\mu
    +O(\mu^2),
\end{aligned}
\label{eq:S3_small_mu_gap12}
\end{equation}
with
\begin{equation}
    \Delta E_{12}'(0)>0.
\end{equation}
The threshold slope separates the spectral and projected-coupling contributions as
\begin{equation}
    \lambda_c'(0)
    =
    \lambda_c(0)
    \left[
    \frac{\Delta E_{12}'(0)}{\Delta E_{12}(0)}
    -
    \frac{\Gamma_{12}'(0)}{\Gamma_{12}(0)}
    \right].
    \label{eq:S3_initial_slope_condition}
\end{equation}
For $L=10$ and $\kappa=0.02$, evaluation of the exact projected matrix elements in Eq.~\eqref{eq:S3_projected_interchain} gives
\begin{equation}
    \lambda_c(\mu)
    =
    0.11
    +
    0.072\,\mu
    +
    O(\mu^2),
    \label{eq:S3_L10_small_mu_lambda}
\end{equation}
so the widening of the $(1,2)$ level separation dominates the initial response. At fixed $\kappa L$, the relative opening of this band-edge detuning scales as
\begin{equation}
    \frac{\Delta E_{12}'(0)}{\Delta E_{12}(0)}
    \simeq
    \frac{20\cosh(\kappa L)}{3(L+1)},
    \label{eq:S3_relative_gap_opening}
\end{equation}
which also explains why this upward trend is most visible for the smaller system.
More generally, a pair-resolved threshold $\lambda_c^{(i,j)}=\Delta E_{ij}/\Gamma_{ij}$ rises with a control parameter whenever the relative growth of its reference detuning exceeds that of its projected activation strength, $\partial_\mu\ln\Delta E_{ij}>\partial_\mu\ln\Gamma_{ij}$. The upward trend is therefore a competition of relative rates, not a special property of the present boundary coupling.

The subsequent decreasing slope of the threshold has a different origin. After the lower-envelope switches $(1,2)\rightarrow(2,3)\rightarrow(4,5)$, the terminal pair $(4,5)$ approaches the isolated-chain exceptional point.
The three activated channels in Eq.~\eqref{eq:S3_L10_sequence} have qualitatively different dependences on the boundary coupling. The band-edge channel $(1,2)$ is regular around $\mu=0$: both $\Delta E_{12}$ and $\Gamma_{12}$ are analytic functions of $\mu$, and the increase of $\Delta E_{12}$ dominates, producing the positive initial slope derived above. The intermediate channel $(2,3)$ is also nonsingular throughout the interval in which it is selected. It becomes activated not because either of its ingredients develops a singularity, but because the smooth ratio
$
\Delta E_{23}(\mu)/\Gamma_{23}(\mu)
$
crosses below the corresponding ratios of the neighboring channels. By contrast, the terminal pair $(4,5)$ is singular because it is the pair that coalesces at the isolated-chain exceptional point $\mu_c^{(0)}$.
To make the latter behavior explicit, let
\begin{equation}
\delta\mu=\mu_c^{(0)}-\mu>0.
\end{equation}
At the isolated-chain exceptional point, the secular equation has a generic double root,
\begin{equation}
F_L(k_c,\mu_c^{(0)})=0,
\qquad
\partial_kF_L(k_c,\mu_c^{(0)})=0.
\end{equation}
The regular channel $(1,2)$ does not require a double-root expansion because both $F_L$ and $\partial_kF_L$ are nonzero on its branches: no reference levels coalesce there, so an ordinary Taylor expansion in $\mu$ is sufficient. A double-root expansion is needed only for the terminal pair that actually approaches the reference EP. Expanding Eq.~\eqref{eq:S3_secular_equation} around that double root gives
\begin{equation}
0=
-(\partial_\mu F_L)_c\delta\mu
+
\frac{1}{2}(\partial_k^2F_L)_c
(k-k_c)^2
+
O\!\left(\delta\mu^{3/2}\right),
\label{eq:S3_double_root_expansion}
\end{equation}
where the subscript $c$ denotes evaluation at $(k_c,\mu_c^{(0)})$. Provided
$(\partial_\mu F_L)_c\neq0$ and $(\partial_k^2F_L)_c\neq0$, the two roots associated with the terminal pair split as
\begin{equation}
k_{4,5}(\mu)
=
k_c
\mp
a(\delta\mu)^{1/2}
+
O(\delta\mu),
\qquad
a>0.
\label{eq:S3_terminal_root_splitting}
\end{equation}
Since $E(k)=2\cos k$ and $\sin k_c\neq0$ for the $L=10$ exceptional point, their reference-energy detuning obeys
\begin{equation}
\Delta E_{45}(\mu)
=
4a|\sin k_c|(\delta\mu)^{1/2}
+
O(\delta\mu).
\label{eq:S3_terminal_detuning}
\end{equation}
The projected activation strength has the opposite singular behavior. As the two reference eigenvectors approach a second-order exceptional point, they become parallel and their biorthogonal self-overlaps vanish. After imposing the normalization in Eq.~\eqref{eq:S3_biorthogonality}, the corresponding rank-one projectors, and hence the projected inter-chain matrix elements entering Eq.~\eqref{eq:S3_Gamma_pq}, generically diverge inversely with the root separation. Therefore,
\begin{equation}
\Gamma_{45}(\mu)
=
\gamma_{45}
(\delta\mu)^{-1/2}
+
O(1),
\qquad
\gamma_{45}>0,
\label{eq:S3_terminal_activation_divergence}
\end{equation}
unless an additional symmetry enforces a cancellation. Combining Eqs.~\eqref{eq:S3_terminal_detuning} and \eqref{eq:S3_terminal_activation_divergence} in the pair-resolved threshold gives
\begin{equation}
\lambda_c^{(4,5)}(\mu)
=
\frac{\Delta E_{45}(\mu)}
{\Gamma_{45}(\mu)}
=
C_{45}
\left[\mu_c^{(0)}-\mu\right]
+
O\!\left[(\mu_c^{(0)}-\mu)^{3/2}\right],
\label{eq:S3_terminal_linear_collapse}
\end{equation}
with $C_{45}>0$.
The nonmonotonic threshold therefore results from a competition between three distinct channel behaviors. The regular band-edge channel $(1,2)$ initially moves upward because its reference detuning widens. The regular intermediate channel $(2,3)$ temporarily becomes the lowest threshold through an ordinary crossing of smooth pair-resolved curves. The pair $(4,5)$ is singular precisely because it is the pair satisfying $F_L(k_c,\mu_c^{(0)})=\partial_kF_L(k_c,\mu_c^{(0)})=0$, which defines the isolated-chain EP. All other candidate pairs remain spectrally isolated as $\mu\to\mu_c^{(0)}$, so their biorthogonal projectors and $\Gamma_{ij}$ remain finite. The terminal channel $(4,5)$ therefore takes over because it alone develops both a vanishing detuning and a divergent projected activation strength. The boundary coupling thus changes not only the numerical value of the threshold, but also the physical mechanism responsible for the first instability.
This mechanism is not restricted to $L=10$. Whenever different candidate pairs respond differently to a control parameter, their pair-resolved thresholds can exchange order and generate nonmonotonic or piecewise behavior in the physical lower envelope. In particular, a regular channel may control the transition far from a reference exceptional point, while a terminal channel with square-root spectral closing and enhanced biorthogonal coupling necessarily dominates sufficiently close to it.

The comparison between Figs.~\ref{fig:S3_ReIm_lambda_L10} and \ref{fig:S3_ReIm_lambda_L30} shows that continuous spectral deformation alone does not determine the number of activated-channel crossovers. For $L=10$, the first coalescing branches change twice, producing the sequence $(1,2)\rightarrow(2,3)\rightarrow(4,5)$. For $L=30$, the band-edge channel remains lowest until a single direct handoff to the terminal pair $(14,15)$ near the isolated-chain threshold. Thus, the number and location of the crossovers are determined by the global lower envelope of Eq.~\eqref{eq:S3_global_threshold}, not by a fixed relabeling rule or by the motion of the reference levels alone.
More generally, activated-channel switching provides a generic route to nonmonotonic real-to-complex thresholds. If different candidate pairs respond qualitatively differently to a control parameter—for example, one channel is hardened by increasing spectral detuning, whereas another is softened by decreasing detuning or by enhanced projected coupling—their pair-resolved thresholds can acquire opposite slopes. The physical threshold, being their lower envelope, then changes its trend when the minimizing pair switches, even if every individual pair-resolved branch remains smooth and monotonic. Proximity of the terminal pair to a reference exceptional point, as in the present example, makes this reversal particularly robust but is not essential: nonmonotonicity generally reflects a change in the microscopic pair responsible for the first instability, rather than a phase transition or a nonanalytic change of the underlying Hamiltonian. 

The general physical implication of this reference-EP proximity, including the linear threshold collapse and the role of the divergent biorthogonal projector, is summarized in End Matter Sec.~IV and Eq.~(29).

\section{Topological-edge-channel threshold in the locally activated SSH chain}
\label{sec:S5_SSH_edge_scaling}

The main text introduced the OBC SSH chain with reflection-related local gain and loss [Eqs.~(20) and (21)]. Its pair-resolved lower envelope switches from the topological edge pair to a bulk pair even though the reference Hamiltonian and its topological invariant remain fixed. For the edge-controlled regime, the main text reported $\lambda_c^{\rm SSH}(m)\simeq|w|r^{L-2m+2}$ [Eq.~(22)]. This section derives the two exponential factors behind that result. There are $L$ unit cells, the intracell and intercell hoppings are $v$ and $w$, and
\begin{equation}
r=\left|\frac{v}{w}\right|<1.
\label{eq:S5_r_definition}
\end{equation}
In the well-localized regime, a convenient normalized basis for the left and right edge states is
\begin{equation}
\begin{aligned}
|L_{\rm e}\rangle
&=\mathcal N\sum_{n=1}^{L}
\left(-\frac{v}{w}\right)^{n-1}|A_n\rangle,\\
|R_{\rm e}\rangle
&=\mathcal N\sum_{n=1}^{L}
\left(-\frac{v}{w}\right)^{L-n}|B_n\rangle,
\end{aligned}
\qquad
\mathcal N^2=\frac{1-r^2}{1-r^{2L}}.
\label{eq:S5_localized_edge_states}
\end{equation}
These states occupy opposite sublattices and overlap only through the finite chain. Projecting $\mathcal H_0$ onto their span gives
\begin{equation}
\langle L_{\rm e}|\mathcal H_0|R_{\rm e}\rangle
=-\mathcal N^2w\left(-\frac{v}{w}\right)^L,
\label{eq:S5_edge_hybridization}
\end{equation}
up to corrections beyond the leading edge-state approximation. The opposite-parity eigenstates
$|\psi_\pm\rangle\simeq(|L_{\rm e}\rangle\pm|R_{\rm e}\rangle)/\sqrt2$
therefore have the finite-size splitting
\begin{equation}
\Delta E_{\rm edge}
=|E_+-E_-|
\simeq2\mathcal N^2|w|r^L
\propto r^L.
\label{eq:S5_edge_splitting}
\end{equation}

The activation operator is local,
\begin{equation}
\mathcal V_m=-i|A_m\rangle\langle A_m|+i|B_{L+1-m}\rangle\langle B_{L+1-m}|.
\label{eq:S5_SSH_defect_local}
\end{equation}
At the two reflection-related defect sites, Eq.~\eqref{eq:S5_localized_edge_states} gives equal edge-state probabilities
\begin{equation}
|\langle A_m|L_{\rm e}\rangle|^2
=|\langle B_{L+1-m}|R_{\rm e}\rangle|^2
=\mathcal N^2r^{2m-2}.
\label{eq:S5_edge_probabilities}
\end{equation}
Consequently, the diagonal difference vanishes in the parity basis and its off-diagonal projected element is
\begin{equation}
A_{+-}=\langle\psi_+|\mathcal V_m|\psi_-\rangle
\simeq i\mathcal N^2r^{2m-2},
\qquad
|A_{+-}|\propto r^{2m-2}.
\label{eq:S5_edge_projected_coupling}
\end{equation}
For this Hermitian reference problem with balanced gain and loss, the edge-pair EP threshold is $\lambda_{c,\rm edge}=\Delta E_{\rm edge}/(2|A_{+-}|)$. Substitution of Eqs.~\eqref{eq:S5_edge_splitting} and \eqref{eq:S5_edge_projected_coupling} cancels the common normalization factor and yields
\begin{equation}
\boxed{
\lambda_{c,\rm edge}^{\rm SSH}(m)
\simeq |w|r^{L-2m+2}.}
\label{eq:S5_SSH_edge_threshold}
\end{equation}
Thus the apparently non-local reach of the topological edge channel does not arise from a non-local Hamiltonian or activation operator. It is the ratio of the exponentially small edge splitting, $r^L$, to the exponentially small local defect projection, $r^{2m-2}$. The edge pair remains activated until Eq.~\eqref{eq:S5_SSH_edge_threshold} rises above the lowest bulk-pair threshold in the global lower envelope.

\end{document}